\documentclass[english,aps,prb,superscriptaddress,twocolumn]{revtex4-2}
\usepackage{babel}
\usepackage{amsmath,mathtools}
\usepackage{amssymb}
\usepackage{graphicx}
\begin{document}
\title{ System size dependent topological zero modes in coupled topolectrical chains }
\author{S M Rafi-Ul-Islam }
\email{e0021595@u.nus.edu}
\selectlanguage{english}%
\affiliation{Department of Electrical and Computer Engineering, National University of Singapore, Singapore}
\author{Zhuo Bin Siu}
\email{elesiuz@nus.edu.sg}
\selectlanguage{english}%
\affiliation{Department of Electrical and Computer Engineering, National University of Singapore, Singapore}
\author{Haydar Sahin}
\email{sahinhaydar@u.nus.edu}
\selectlanguage{english}%
\affiliation{Department of Electrical and Computer Engineering, National University of Singapore, Singapore}
\affiliation{Institute of High Performance Computing, A*STAR, Singapore}
\author{Ching Hua Lee}
\email{phylch@nus.edu.sg}
\selectlanguage{english}%
\affiliation{Department of Physics, National University of Singapore, Singapore}
\author{Mansoor B. A. Jalil}
\email{elembaj@nus.edu.sg}
\selectlanguage{english}%
\affiliation{Department of Electrical and Computer Engineering, National University of Singapore, Singapore}
\begin{abstract}
In this paper, we demonstrate the emergence and disappearance of topological zero modes (TZMs) in a coupled topolectrical (TE)  circuit lattice. Specifically, we consider non-Hermitian TE chains in which TZMs do not occur in the individual uncoupled chains, but emerge when these chains are coupled by inter-chain capacitors. The coupled system hosts TZMs which show size-dependent behaviours and vanish beyond a certain critical size. In addition, the emergence or disappearance of the TZMs in the open boundary condition (OBC) spectra for a given size of the coupled system can be controlled by modulating the signs of its inverse decay length. Analytically, trivial and non-trivial phases of the coupled system can be distinguished by the differing ranks of their corresponding Laplacian matrix. The TE circuit framework enables the physical detection of the TZMs via electrical impedance measurements. Our work establishes the conditions for inducing TZMs and modulating their behavior in coupled TE chains.
\end{abstract}
\maketitle
\section{Introduction}
One of the most interesting phenomena in condensed matter system is the discovery of novel topologically protected states such as edge states \cite{hafezi2013imaging,obana2019topological,st2017lasing,rafi2021topological,gao2020observation}, zero-energy modes \cite{ganeshan2013topological,kempkes2019robust,ryu2002topological}, corner modes \cite{imhof2018topolectrical,sahin2022interfacial,banerjee2020coupling,wu2021topological} and, hinge modes \cite{kunst2018lattice,benalcazar2017electric,ghorashi2019second} in various Hermitian systems. These topologically protected states constitute a new basis for diverse physical phenomena \cite{jin2018robustness,hafezi2013imaging,fujita2011gauge,tan2020yang} and various related applications \cite{wu2017applications,vobornik2011magnetic,kesselring2018boundaries} because of their many interesting characteristics such as their robustness  against system disorders and perturbations. The topologically non-trivial phases in such Hermitian systems are characterized by Bloch wave-vectors that respect the usual bulk-boundary correspondence (BBC), and their band spectra under open boundary conditions (OBC) in the limit of infinite system size are identical to that  under periodic boundary conditions (PBC). The introduction of non-Hermiticity, for example in the form of non-reciprocal couplings \cite{helbig2020generalized,qiu2019unidirectional,santos2014non}, may result in the emergence of drastic differences between the OBC and PBC spectra and the breakdown of the usual BBC and  of the Bloch theorem \cite{rafi2021non,li2022non,yang2022designing,zeuner2015observation,tai2022zoology,li2021quantized,gu2016holographic,leykam2017edge,gong2018topological,el2018non,yokomizo2019non}. Several non-Hermitian systems have been realized in various platforms ranging from topolectrical \cite{rafi2020topoelectrical,zhang2022anomalous,rafi2021unconventional,lenggenhager2021electric,rafi2020anti,stegmaier2021topological,rafi2020realization,lee2018topolectrical,rafi2020anti,hofmann2020reciprocal,rafi2021unconventional,kotwal2019active,helbig2020generalized}, photonics \cite{zhu2020photonic,song2020two,xiao2020non}, and acoustic \cite{zhang2020non,gao2020anomalous} systems,  as well as superconductors \cite{wang2021majorana,zhou2020non,cao2021universal} and metamaterials \cite{schomerus2020nonreciprocal,zhou2020non,ghatak2020observation}. In such systems, the wavefunctions are localized in the vicinity of  the system boundaries under OBC, a phenomenon known as  the non-Hermitian skin effect (NHSE) \cite{rafi2021unconventional,longhi2019probing,rafi2021critical,okuma2020topological,li2020critical,song2019non,rafi2021unconventional,kawabata2020higher,yao2018edge,li2020topological}.

Interestingly, the introduction of coupling between two non-Hermitian chains with dissimilar degree of non-Hermiticity fundamentally alters the topological character of the coupled system \cite{rafi2022critical}. It was recently found that coupled non-Hermitian chains exhibit scale-free exponential wavefunctions \cite{li2021impurity} and the critical NHSE (CNHSE) \cite{li2020critical,zhang2021experimental,liu2020helical,sahin2022unconventional,yokomizo2021scaling,liu2020helical,rafi2021critical,liu2020helical}, in which the wavefunctions and eigenvalues experience a discontinuous transition as the system size is increased beyond some critical point. However, much remains to be understood regarding the evolution of the topological edge modes in such coupled non-Hermitian systems including their dependence on system size.\\
In the following, we demonstrate via numerical and analytical results the emergence  of topologically protected zero modes (TZM) in a coupled non-Hermitian topolectrical (TE) circuit chain consisting of inductors, capacitors, and op-amps. We found that, under some specific combination of inter-chain hopping and non-Hermiticity parameters, TZMs can emerge in the coupled system even when the individual chains do not  host any TZMs. Conversely, when the individual chains are tuned to reside in the non-trivial regimes, a finite inter-chain coupling can result in size-dependent zero modes that persist up to critical system size and then vanish abruptly upon further increase in the system size. Analytically, the emergence and disappearance of the TZMs via inter-chain coupling can be predicted by evaluating the rank of a matrix constructed from the eigenvectors of the surrogate Hamiltonian of the circuit. Furthermore, in the physical TE circuit realization, the presences of TZMs can be detected by impedance spectral measurements. In summary, we showed the emergence and modulation of TZMs in coupled TE systems by varying either the inter-chain coupling strengths or the system size.

\begin{figure*}[ht!]
\centering
\includegraphics[width=0.8\textwidth]{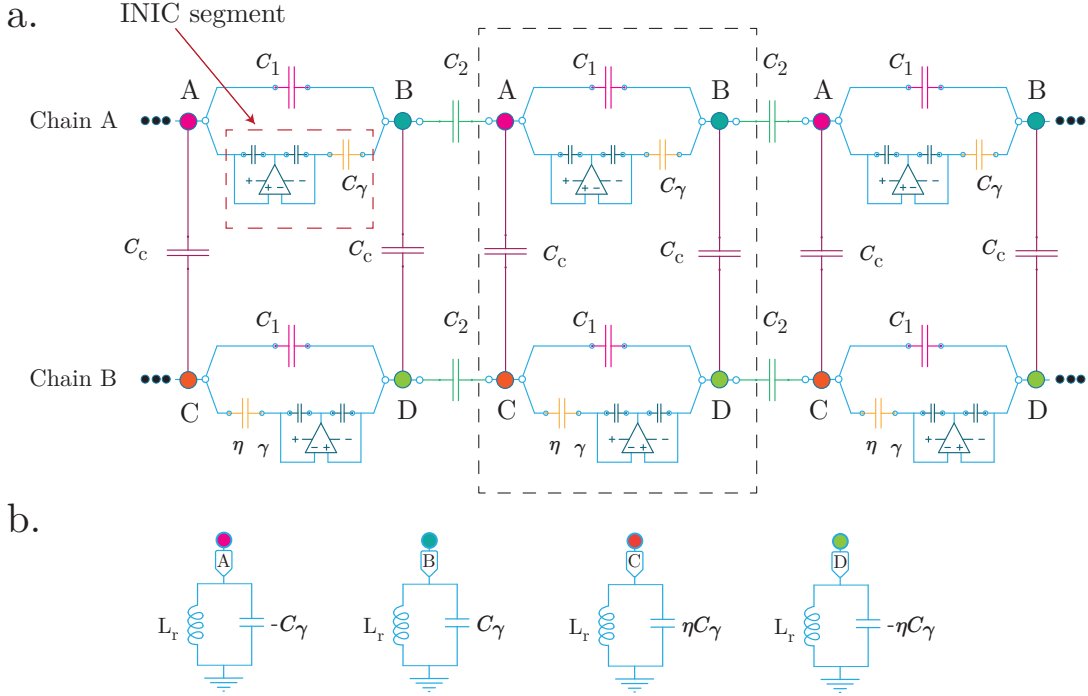}
\caption{ Schematic of a pair of coupled non-Hermitian TE chains. a. The coupled TE circuit comprises of two non-Hermitian SSH chains with opposite signs of the non-Hermiticity capacitances $C_\gamma$. Within each chain, the intra unit cell coupling is directionally asymmetric; in the top (bottom) chain, the couplings in the two directions are given by $C_1 \pm C_\gamma (C_1 \pm \eta C_\gamma)$, respectively. The coupling asymmetry is implemented by negative impedance converter with current inversion (INIC). The two chains are coupled via a inter-chain capacitor $C_c$. b. Grounding mechanism for the different types of nodes. The couplings to the ground are chosen such that a single resonant frequency of $f_r=(2\pi \sqrt{L_r(C_1+C_2+C_c)})^{-1}$ is established via tuning the common inductance $L_g$, so that the onsite potential is zero at every node.}
\label{gFig1}
\end{figure*}

\section*{Results}
\subsection{Construction of coupled TE chains that host TZMs}
The pivotal step in inducing TZMs  is to construct a TE circuit lattice that emulates the electronic band structure of a coupled non-Hermitian system having skin modes of  dissimilar inverse decay lengths. (The inverse decay length is the imaginary part of the complex wavevector of the OBC mode.)  To do so, we consider the TE circuit array shown in Fig. \ref{gFig1} formed by connecting two non-Hermitian chains (top and bottom rows of Fig. \ref{gFig1}a) by an inter-chain coupling capacitance $C_c$. In the absence of the inter-chain coupling, each of the two chains are analogous to the non-Hermitian SSH model with asymmetric directional couplings between two neighboring nodes in their unit cell. These are given by ($C_1 \pm C_\gamma$) in the two directions for the top chain, and by ($C_1 \pm \eta C_\gamma$) for the bottom chain. For both chains, there is a reciprocal  coupling of $C_2$ between adjacent unit cells (i.e., the coupling coefficient in the direction from a site to its neighbour is the same as that in the direction from the neighbour to the site.) The coupling asymmetry in the intra-chain segment is realized in the practical circuit via negative impedance converters with current inversion (INICs)\cite{rafi2021non}. The non-Hermiticity parameter $C_\gamma$ can be modulated so that the  eigenmodes of the upper and lower chains would have different inverse decay lengths   when $\eta \neq 1$. When the parameter $\eta$ is set to $-1$, the two chains are in an antisymmetric configuration where the sum of the inverse decay lengths of the two chains is zero.
\begin{figure*}[ht!]
\centering
\includegraphics[width=0.8\textwidth]{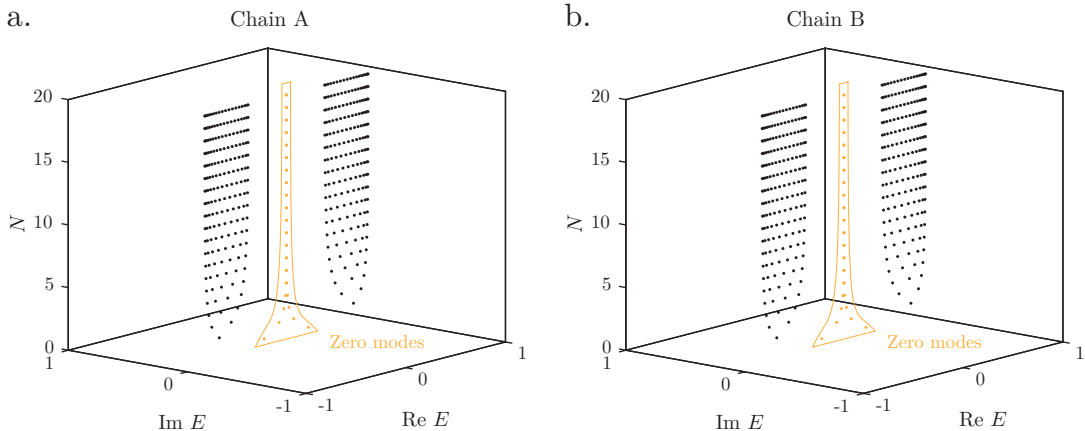}
\caption{ The OBC spectra as a functions of the system size for the individual uncoupled $A$ and $B$ chains for later comparison with the coupled chain system. a), b) The OBC eigenenergy distributions for chains $A$ and $B$, respectively,  at various system sizes $N$. Other common parameters:  $\eta =1$, $C_1 = 0.83$, $C_2 = 0.6$, and  $C_\gamma = 0.8$. The topological zero modes are evident at all system lengths, unlike the coupled system which we shall describe later.  }
\label{gFign10}
\end{figure*}
The corresponding Laplacian for the circuit in Fig. \ref{gFig1}a (multiplied by $1/(i\omega)$) can be expressed as 
 \begin{equation}
(i \omega)^{-1}L(k, \omega_{res}) = \begin{pmatrix}
H_a (k, \omega_{res}) & C_c \mathcal{I}_{2\times 2} \\
C_c \mathcal{I}_{2\times 2} & H_b (k, \omega_{res})
\end{pmatrix},
\label{eq1}
\end{equation}
where 
\begin{equation}
\begin{aligned}
&H_a (k, \omega_{res})= \\ &=\begin{pmatrix}
(C_1+C_2+C_c-\frac{1}{\omega^2 L_r}) & C_1+C_{\gamma}+C_2 \exp(-ik)\\
C_1-C_{\gamma}+C_2 \exp(ik) & (C_1+C_2+C_c-\frac{1}{\omega^2 L_r})
\end{pmatrix}
\end{aligned}
\end{equation}
 and  
\begin{equation}
\begin{aligned}
&H_b (k, \omega_{res})=\\ &=\begin{pmatrix}
(C_1+C_2+C_c-\frac{1}{\omega^2 L_r}) & C_1- \eta C_{\gamma}+C_2 \exp(-ik)\\
C_1+ \eta C_{\gamma}+C_2 \exp(ik) & (C_1+C_2+C_c-\frac{1}{\omega^2 L_r})
\end{pmatrix}.
\end{aligned}
\end{equation} 
 We introduce the corresponding surrogate Hamiltonian to the Laplacian and the non-Bloch factor $\beta = e^{i k}$ as $H(\beta) \equiv (i\omega)^{-1}L(k = -i\ln\beta, \omega)$ where $k$ can now take or imaginary on complex values  under OBC. We will show that Eq. \ref{eq1} represents a coupled TE system that exhibits size-dependent non-trivial boundary states. 

\subsection{System size-dependent topological zero modes}

To highlight the features of the coupled chain system, we first study the OBC of the individual uncoupled chains for comparison. The OBC admittance distributions are plotted at varying system size $N$ in Fig. \ref{gFign10} for the individual uncoupled chains $A$ and $B$ at $\eta=1$, which implies that the coupling strength for hoppings from left to right in the first chain is equal to the coupling strength for the hopping in the reversed in the second chain, and vice-versa, i.e., ``anti-symmetric coupled chains''.  Following the properties of generic non-Hermitian SSH chains \cite{kunst2018biorthogonal,lieu2018topological},  both individual chains host TZMs for the parameter range of  $-\sqrt{C_2^2+C_\gamma^2}<C_1<\sqrt{C_2^2+C_\gamma^2}$. In the limit of large system size, the TZMs of the uncoupled chains are degenerate at zero energy, and their behaviour are independent of the system size (see Fig. \ref{gFign10}a, b). However for small system size (i.e. of the order of $N \approx 4$), the TZMs become non-degenerate and are no longer pinned at zero energy (see Fig. \ref{gFign10}a, b). (The observed behaviour of the TZMs of uncoupled chains are discussed in more detail in the Appendix).  
 
\begin{figure*}[ht!]
\centering
\includegraphics[width=\textwidth]{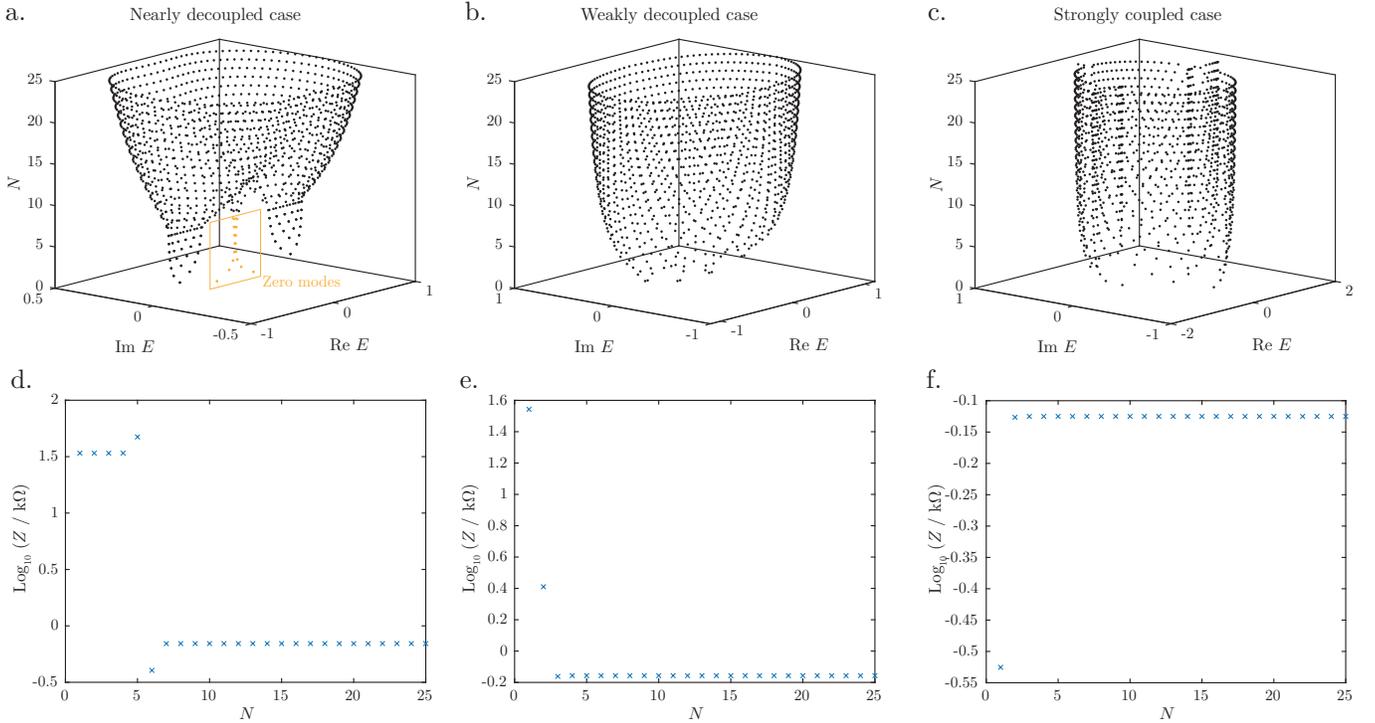}
\caption{Anomalous disappearance of topological modes as system size is varied. Demonstration of the effect of system size and the inter-chain coupling $C_c$ on the eigenenergy distribution for coupled TE chains. a)-c) The OBC dispersion relations with respect to the system size at $\eta=1$ for $C_c = 10^{-7}, 0.005, 0.3$ respectively. Zero-modes exist at small $C_c$ and small $N$. d)-f) Comparison of the impedance  $Z$ as functions of $N$ for three values of $C_c = 10^{-7}, 0.005, 0.3$ respectively at $\eta=1$. High and low impedance readouts signify the presence and absence of the topological zero-modes, respectively. Other parameters:   $C_1 = 0.83$, $C_2 = 0.6$ and  $C_\gamma = 0.8$. The zero modes vanish at strong inter-chain coupling and / or long chains. }
\label{gFign11}
\end{figure*}

Next, we plot the variation of OBC eigenenergy spectra as a function of $N$ for various values of the inter-chain couplings strengths ($C_c$) corresponding to the nearly decoupled ($C_c = 10^{-7}$), weakly coupled ($C_c = 0.005$), and strongly coupled ($C_c = 0.3$) systems. We set $\eta=1$, which implies that the coupling strength for hoppings from left to right in the first chain is equal to the coupling strength for the hopping from right to left in the second chain, and vice-versa, i.e., ``anti-symmetric coupled chains''.  

In nearly decoupled systems with very small $C_c$ values (on the order of $10^{-7}$), the coupled systems host TZMs as long as the OBC spectra lie on the real axis. These real energy spectra with TZMs in the coupled systems occur at relatively small system sizes (see Fig. \ref{gFign11}a). In other words, the TZMs survive for system sizes less than some critical $N= N_{\mathrm{critical}}$. When $N> N_{\mathrm{critical}}$, the OBC eigen spectra expand into the complex plane and do not exhibit any TZMs (the four-fold degenerate TZMs split and move away from the real axis). Therefore, in the nearly decoupled TE chains, the TZMs vanish beyond a critical system size \cite{rafi2021critical}, i.e. they exhibit the non-Hermitian skin effect (CNHSE) (see Fig. \ref{gFign11}a). The emergence and disappearance of these peculiar size-dependent TZMs in the coupled TE chains can be characterized by impedance measurements between the two leftmost nodes of Chain A as shown in Fig. \ref{gFign11}d. Here, the presence or absence of the TZMs is distinguished by high and low impedance readouts, respectively. Furthermore, a sharp transition between the high and low impedance states occurs marks the critical system size corresponding to the CNHSE. 

As the magnitude of $C_c$ increases, the hybridization between the two chains becomes more prominent and the characteristics of the coupled chains deviate drastically from that of the uncoupled chains. For instance, even in  the case of  weakly coupled chains ($C_c=0.005$), the CNHSE transition occurs at a much smaller system size ($N \approx 1$) for our choice of model parameters (see Fig. \ref{gFign11}b). No TZMs exist for $N> N_{\mathrm{critical}}$, resulting in the low impedance readouts in Fig. \ref{gFign11}e. When the inter-chain coupling $C_c$ takes on a significant value ( i.e., $C_c=0.3$) which is  on the order of the other parameters, the strong hybridization between two chains prevents the emergence of TZMs and the OBC energy spectra become almost independent of system size (see Fig. \ref{gFign11}c). The absence of the TZMs translates into very small impedance measurements for all system size (see Fig. \ref{gFign11}f). 
\begin{figure*}[ht!]
\centering
\includegraphics[width=0.70\textwidth]{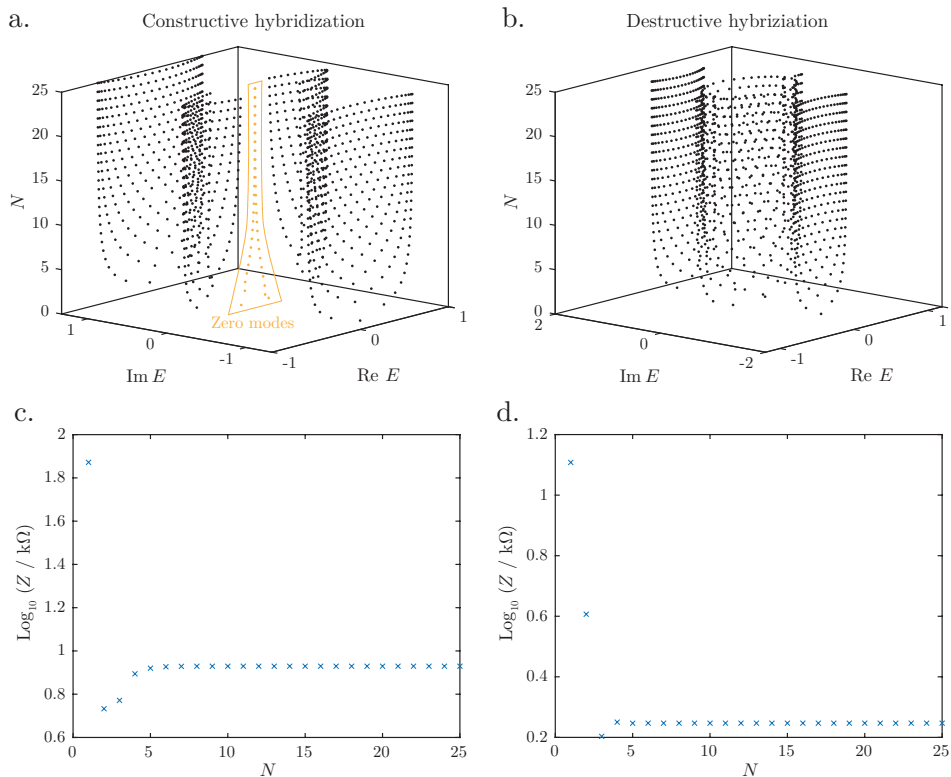}
\caption{Impedance and zero-mode evolution under OBC for various hybridization scenarios between TE chains. a),b) Evolution of TZM states as a function of system size ($N$) for constructive (i.e., $\eta=-2$) and destructive (i.e., $\eta=2$) hybridization, respectively. c),d) The variation of  impedance $Z$  with respect to system size ($N$) for constructive and destructive hybridization respectively.
Other common parameters:   $C_1 = 0.83$, $C_2 = 0.6$, $C_c = 0.2$, and $C_\gamma = 0.8$. Note that for this set of parameters, TZMs  exist only under condition of constructive hybridization. }
\label{gFign12}
\end{figure*}
Next, we plot the OBC energy spectra and impedance readout with respect to $N$ for the general case of  $\eta \neq 1$, in which the coupling in the two chains are no longer exactly anti-symmetric. For the case where $\eta<0$, the $\ln|\beta|$ values of the eigenstates localized in the two chains $A$ and $B$  (i.e., $\ln|\beta|_{A-chain}=\frac{C_1+C_\gamma}{C_1-C_\gamma}$ and $\ln|\beta|_{B-chain}=\frac{C_1+\eta C_\gamma}{C_1-\eta C_\gamma}$) have the same sign. In this configuration, which we describe as exhibiting ``constructive hybridization'', (see \cite{rafi2021critical}), there are well-defined TZMs in the OBC spectra (see Fig. \ref{gFign12}a). Furthermore, the TZMs exist for all values of system size $N$, and thus the system no longer exhibits CNHSE. However, for the case where $\eta$ has a positive value ($\eta \ne 1$), i.e., the $ln |\beta|$ has opposite signs for the states localized in the two chains, the configuration corresponds to "destructive hybridization'' and the TZMs vanish (see Fig. \ref{gFign12}b). As before, the presence and absence of TZMs under constructive and destructive hybridization can be distinguished by the relatively large and small impedance measurements, respectively,  for a given $N$ (see Fig. \ref{gFign12}c-d).
\begin{figure*}[ht!]
\centering
\includegraphics[width=0.70\textwidth]{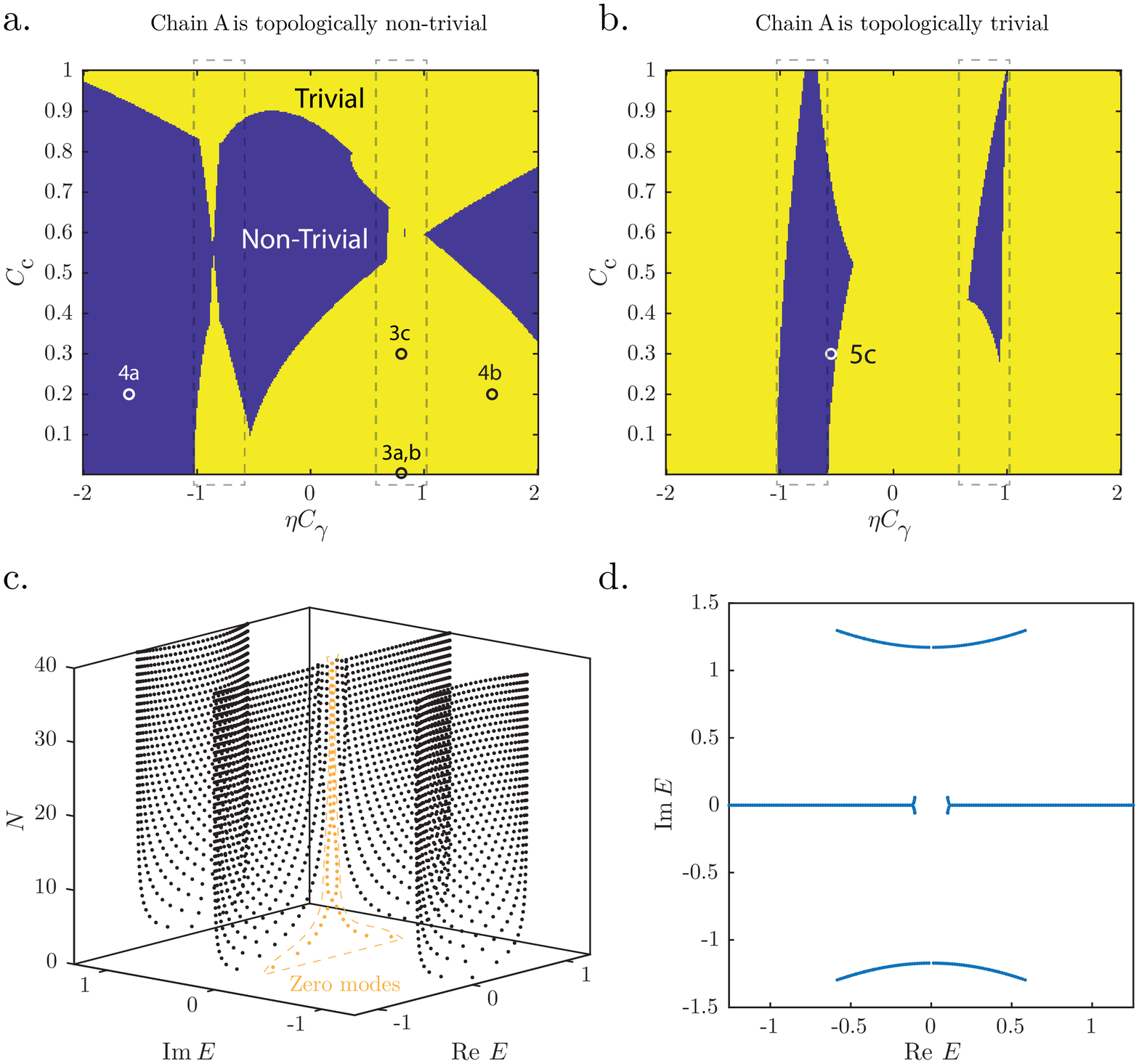}
\caption{a.  Phase diagram of the coupled system with respect to $C_c$ and $\eta C_\gamma$. The other parameter values are chosen to be the same as those in Figs. \ref{gFign11} and \ref{gFign12} for which the isolated $A$ chain is topologically non-trivial. The label above each circle denotes the OBC spectra in Figs. \ref{gFign11} and \ref{gFign12}, which the corresponding parameter coordinates $(C_c, \eta C_\gamma)$ correspond to. b. Phase diagram of the coupled system over the $C_c$-$\eta C_\gamma$ parameter space for the values of the other parameters as in Fig. \ref{gcomPD}a, except that $C_\gamma$ is now set to 1.6, so that chain $A$ is now topologically trivial. The circle labelled "5c" denotes the parameter pair of $(C_c = 0.3, \eta C_\gamma = -0.55)$ for which the admittance spectra are plotted in panels c and d, respectively. The rectangles with the dotted borders indicate the range of $\eta C_\gamma$ for which chain B is topologically non-trivial. c. The variation of the admittance spectra for $C_\gamma = 1.6, C_c = 0.3, \eta C_\gamma=-0.55$ with the system size $N$. d. The OBC admittance spectrum for the parameter set in c. in the thermodynamic limit (i.e., $N \to \infty$) found from the criteria that the admittance values with the middle magnitudes should be coincident. Note the absence of the TZMs highlighted in yellow in panel c here. These results show that topologically zero modes may emerge or vanish in coupled chain systems regardless of the topological character of the two the constituent uncoupled chains, i.e. TZMs can exist in the coupled system when neither or either or both of the constituent chains are topologically non-trivial. }
\label{gcomPD}
\end{figure*}

A condition for the existence of a TZM in the thermodynamic limit is that the system of (four) linear equations governing the boundary conditions at the boundary where the TZM is localized (to ensure that the wavefunction vanishes at that boundary) is not of full rank \cite{lee2019anatomy}. This fact allows a quick analytical method of determining whether the system is topologically non-trivial and hosts TZMs from the number of linearly independent zero admittance eigenvectors among the first eigenvectors with the four smallest magnitudes of $\beta$. In other words, we evaluate  the rank of the matrix constructed from the column-wise concatenation of these zero-admittance eigenvectors. The Hamiltonian of the system would be deemed non-trivial if the rank of the matrix is less than four. Details of this analytical procedure and the rationale for this criterion are presented in the Appendix.

Fig. \ref{gcomPD}a shows the phase diagram of the system as a function of coupling capacitance $C_c$ and non-Hermitian and asymmetric coupling parameter $\eta C_\gamma$ with the other parameters taking the same values as those in Fig. \ref{gFign11}(a) to (c) and \ref{gFign12}(a),(b). The ranges of $C_c$ and $\eta C_\gamma$ shown encompass those considered in Fig. \ref{gFign11} and \ref{gFign12}, for which the positions in the phase diagram corresponding to the plots in these figures are denoted by circles with the corresponding labels. In particular, Fig. \ref{gcomPD} shows that the parameter set in Fig. \ref{gFign11}(a) corresponds to the non-trivial regime. The absence of TZMs at large $N$ for the parameter set in Fig. \ref{gFign11} can now be ascribed to the topologically trivial nature of the coupled chains even though the uncoupled chains in the figure are themselves topologically non-trivial. In the nearly decoupled case in Fig. \ref{gFign11}a, the topologically non-trivial nature of the uncoupled chains are retained in the coupled chains of short system lengths as shown by the existence of the TZMs. However, when the system size exceeds the critical length corresponding to CNHSE the topologically trivial nature of the coupled chains becomes dominant.

In the cases considered so far, the isolated chain A is always non-trivial. We now set $C_\gamma = 1.6$ so that chain A is now trivial and plot the phase diagram under varying $C_c$ and $\eta C_\gamma$, while retaining the same values for the other parameters, as shown in  Fig. \ref{gcomPD}b. In the region outside the dotted rectangles in Fig. \ref{gcomPD}b , both of the uncoupled chains $A$ and $B$ are  trivial. Interestingly, topologically non-trivial phases can emerge when both of these topologically trivial chains are coupled together, as evidenced by the parameter space in the phase diagram lying outside the dotted rectangles that corresponds to the non-trivial state. As an illustration, we consider the specific case of $C_c = 0.3$ and $\eta C_\gamma = 0.55$, which is denoted by a circle labelled ``5c" in Fig. \ref{gcomPD}b. The admittance spectra for increasing finite lengths of the system are shown in Fig. \ref{gcomPD}c. Fig. \ref{gcomPD}d in turn shows the GBZ admittance spectrum obtained as the set of admittance values where the two middle $\beta$ values, arranged in order of increasing magnitude, coincide. Comparing Fig. \ref{gcomPD}c with \ref{gcomPD}d, it can be seen that the TZMs marked in orange in the former are absent in the latter. This indicates that the states marked in orange are indeed topological states, which are not captured in the GBZ, unlike the remaining states, which are bulk states.   


In conclusion, we show that modulation of the inter-chain $C_c$ and non-Hermitian intra-chain $\eta C_\gamma$ couplings can control the onset or disappearance of topological zero modes (TZMs) in the system. The TZMs in the coupled system exhibits a more varied set of behaviour compared to that of single non-Hermitian chains that have been studied hitherto. Our theoretical results reveal that depending on the specific values of $C_c$, $\eta C_\gamma$ and other capacitive couplings, TZMs can be made to appear or vanish in the coupled system for all possible topological character of the two constituent chains. In other words, TZMs can be absent or present for all scenarios, i.e., when neither, either, or both of the uncoupled chains are topologically non-trivial. The emergence of TZMs in the coupled system when neither of the two constituent chains are topologically non-trivial implies the key role played by the inter-chain coupling in determining the topological character of the coupled system. Additionally, in the converse case where both the consituent uncoupled chains are topologically non-trivial while the coupled system is topologically trivial, one observes a size-dependent effect in which TZMs persist at small inter-chain coupling strengths and for small system size, but disappear when the system size exceeds a critical limit, a phenomenon known as the critical non-Hermitian skin effect (CNHSE). We devise an analytical method to distinguish the trivial and non-trivial phases of the coupled system by considering the rank of the matrix constructed from zero-admittance eigenvectors of the surrogate Hamiltonian. Based on this analytical method, we plot the modified phase diagram of the coupled system over the $C_c$ and $\eta C_\gamma$ parameter space, and elucidate the role of the inter-chain and non-Hermitian couplings in determining the topology of the coupled system. In practice, the trivial and non-trivial phases of the coupled TE chains and their evolution with system size can be distinguished by circuit impedance measurements, with the TZMs being associated with higher impedance readouts. More broadly, our results indicate a practical means to induce TZMs and modulate the topological phase transitions in coupled systems based on the TE circuit platforms.

\section{Appendix}
\subsection{Analytical determination of topological non-triviality} 
The eigenstates of a non-Hermitian system may be localized at its edges under OBC because of the two distinct mechanisms of the non-Hermitian skin effect or because they are topologically non-trivial edge states. The former requires boundaries to be present at both ends of the system, and occurs only at energies at which the two median values of $|\beta|$ match in the thermodynamic limit. This arises from the requirement that the linear superposition of the PBC eigenstates constituting the OBC eigenstate satisfy the boundary conditions that the wavefunction of the OBC eigenstate vanishes at both boundaries simultaneously. (See our earlier paper for the details \cite{rafi2021critical}). 

Another mechanism for the emergence of localized states is when these states are topologically non-trivial. A hallmark of such a topologically non-trivial state is that the system of linear equations for the boundary conditions at the boundary where the state is localized near is not of full rank \cite{lee2019anatomy}. 
Here, we describe a quick approach to determine whether a non-Hermitian system with only nearest-neighbor inter-unit cell coupling  supports topologically non-trivial states near zero energy.  Consider the Hamiltonian for a single uncoupled chain with an arbitrary number of nodes, $2n_d$, in each unit cell represented by 
\begin{equation}
	H(\beta) = \begin{pmatrix} \mathbf{0} & \mathbf{H}_-\beta^{-1} + \mathbf{t}_{-+} \\ \mathbf{H}_+\beta + \mathbf{t}_{+-}  & \mathbf{0} \end{pmatrix}. \label{HbetaGen} 
\end{equation}  
where the quantities in bold are $n_d$ by $n_d$ matrices. We write the time-independent Schr\"{o}digner equation for Eq. \eqref{HbetaGen} as 
\begin{equation}
	H(\beta)|\Psi\rangle = |\Psi\rangle E \label{tise} 
\end{equation}
where $|\Psi\rangle$ is the (right) eigenstate with energy $E$. 
The characteristic polynomial for an eigenenergy of 0, $|H(\beta)|=0$, can be made into a quadratic polynomial in $\beta$. This implies that there are, in general, only two finite values of $\beta$ that will result in $H(\beta)$ having an eigenenergy of zero regardless of the value of $n_d$.  When $n_d > 1$, the system also admits the $\beta$ values of 0 and $\infty$: At $\beta = 0$, ($\infty$) the term containing $\beta^{-1}$ ($\beta$) in Eq. \eqref{HbetaGen} dominates over the remaining terms in Eq. \eqref{tise}, including the $|\Psi\rangle E$ term on the right side of the equal sign and Eq. \eqref{tise} effectively becomes $H_+|\Psi\rangle = 0$ ($H_-|\Psi\rangle=0$). There are $n_d-1$ eigenvectors for $\beta=0$, and another $n_d-1$ eigenvectors for $\beta=\infty$. Denoting the $i$th solution of $H_\pm |\Psi\rangle = 0$  as $|\Psi_{\pm;0}^i\rangle$, the $(n_d-1)$ $|\Psi_{\pm;0}^i\rangle$s for each of the two signs of $\pm$ in the subscript constitute eigenvectors with the effective values of $\ln|\beta|=\pm\infty$ that are localized at the right (left) edges of the system. 
For the ease of explanation, let us now consider the simplest example of Eq. \ref{HbetaGen} where all the bolded quantities are scalars, i.e., the single uncoupled SSH chain described by the Hamiltonian 
\begin{equation} 
	H = \begin{pmatrix} 0 & c_{12} + d_{12} / \beta \\ c_{21} + d_{21}\beta & 0 \end{pmatrix} \label{hamSSH1} 
\end{equation}
where $c_{12}$ and $c_{21}$ denote the intra-unit cell coupling, and $d_{12}$ and $d_{21}$ the inter-unit cell coupling ($c_{ij}$ and $d_{ij}$ are non-zero parameters). It can be easily found that the eigenstates and $\beta$ values of Eq. \ref{hamSSH1} at $E=0$ are 
\begin{align}
	 (1, 0)^\mathrm{T} &: \beta = 0, \beta = -c_{21}/d_{21}; \nonumber \\ 
	(0, 1)^\mathrm{T} &: \beta = \infty, \beta = -d_{12}/c_{12} \label{e0egv}. 
\end{align}
We assume for the moment that $|c_{21}/d_{21}| < 1$, and consider a semi-infinite system that spans from $x=1,2, ..., \infty$. In this case, the system can host an edge state $\psi_{\mathrm{left}}(x)$ satisfying the boundary condition that $\psi_{\mathrm{left}}(x = 0) = 0$ given by 
\begin{equation}
	\psi_{\mathrm{left}}(x) = \begin{pmatrix} 1 \\ 0 \end{pmatrix} (\delta_{x,0} - (-c_{21}/d_{21})^x ) \label{leftSemiInf}
\end{equation}
where $x$ in an integer (because the Hamiltonian describes a lattice system), and the $\delta_{x,0}$ is the Kronecker (not Dirac) delta due to the localization of the $\beta=0$ state on the left edge of the system. 
Note that Eq. \eqref{leftSemiInf} is not admissible as an eigenstate of a system with a finite length that has both a left and a right edge, as can be seen from the following: Denoting the length of the finite system as $N$ so that that $x=1,2,...,N$ we see substituting $x\rightarrow (N+1)$ into Eq. \eqref{leftSemiInf} does not satisfy the requirement that the wavefunction vanishes at $x=(N+1)$ because $\psi_{\mathrm{left}}(N+1) = (1, 0)^{\mathrm{T}} (-(c_{21}/d_{21}))^{N+1})$. Moreover, this remaining bit cannot be canceled off by any linear combination of the remaining two eigenvectors, which are proportional to $(0,1)^{\mathrm{T}}$. The formation of edge states that will satisfy the boundary conditions at both edges in finite-length systems therefore requires the energy to be shifted slightly away from $E = 0$ so that the $\beta$ values and eigenvectors in Eq. \eqref{e0egv} now become 
\begin{equation}
\begin{aligned}
	&\begin{pmatrix} 1 \\ 0 \end{pmatrix},\beta = 0; \begin{pmatrix} 1 \\ \delta a_1 \end{pmatrix}, \beta_1 = -c_{21}/d_{21} + \delta b_1;\\ &\begin{pmatrix} \delta a_2 \\ 1 \end{pmatrix},\beta_2 = -d_{12}/c_{12} + \delta b_2; \begin{pmatrix} 0 \\ 1 \end{pmatrix},\beta=\infty
\end{aligned}
\end{equation} 
where the $\delta a_i$s and $\delta b_i$s are small shifts due to the shift in the energy. The presence of the $\delta a_i$s in the eigenspinors of the terms with finite $\beta$s now allow the terms to cancel off one another at both boundaries to satisfy the boundary conditions at both edges. For an edge state localized at the left edge, the resultant (unnormalized) wavefunction can be written as 
\begin{equation}
\begin{aligned}
	\psi_{\mathrm{finite}}(x) =&\begin{pmatrix} 1 \\ 0 \end{pmatrix}\delta_{x,0} + c_1 \begin{pmatrix} 1 \\ \delta a_1 \end{pmatrix} (\beta_1)^x  
\\ &+ d_2 \begin{pmatrix} \delta a_2 \\ 1 \end{pmatrix}(\beta_2)^x  + d_3 \begin{pmatrix} 0 \\ 1 \end{pmatrix} \delta_{x,N+1} \label{psiFinLeft}
\end{aligned}
\end{equation}
where $c_1 \rightarrow -1, \beta_1 \rightarrow (-c_{21}/d_{21}), \beta_2 \rightarrow (-d_{12}/c_{12})$ and $\delta a_1, \delta a_2, d_2, d_3 \rightarrow 0 $ as $N \rightarrow \infty$. The finite length can in this case be interpreted as a perturbation to the ``ideal case'' of the semi-infinite system.
 
Note that for Eq. \eqref{psiFinLeft} to describe a state localized at the left edge, we require $|\beta_1| < |\beta_2|$ so that at the left edge at $x=0$, the $(\delta a_2,1)^{\mathrm{T}}$ state has a miniscule weight compared to the $(1,a_1)^{\mathrm{T}}$ state and the latter is largely cancelled off by the $(1, 0)^{\mathrm{T}}$ state with $\beta=0$, which is confined to only the left edge. A similar consideration for the edge state localized on the right edge of the system would also lead to the requirement for $|\beta_1| < |\beta_2|$ in order for such an edge state to be formed. 

There is another crucial difference between the semi-infinite length system, which has only a single edge, and the finite length system, which has two edges. In the latter, there are no restrictions on whether $|\beta_1|$ or $|\beta_2|$ are required to be smaller or larger than 1 for edge states to exist in the system. In the semi-infinite system extending from $x=[0,\infty)$, it is strictly required that the finite $|\beta|<1$ so that the wavefunction amplitude remains bounded as it extends to infinity. In contrast, in a system of finite length, we saw in Eq. \eqref{psiFinLeft} that in general, a wavefunction satisfying the OBC at both the left and right boundaries requires the simultaneous  combination of the $|\beta|$ values that may be bigger or smaller than one. The finite length of the system ensures that the resulting wavefunction remains bounded to finite values. 
The simple SSH model discussed above suggests the following criteria for quickly establishing whether topological edge states can exist near $E=0$ in more general systems, including the 4-by-4 Hamiltonian considered in the Eq. \eqref{HbetaGen} of the coupled SSH chain system,  without having to perform the far more intensive calculation of the topological invariant of the system: First, solve for the $\beta$ values and their corresponding eigenvectors of the $2n_d$ by $2n_d$ Hamiltonian at the eigenvalue of $E=0$. Arrange the $2n_d$ $\beta$ values (including $\beta=0$ and $\beta=\infty$) in ascending order of $|\beta|$. If the rank of the square matrix formed by the column-wise concatenation of the first  (last) $n_d$ eigenvectors is less than $n_d$, the system hosts a topological edge state on the left (right) edge.  In contrast, if the rank is $n_d$, there can be no edge states. This can be seen from the definition of the rank of a matrix as the number of linearly independent vectors in the matrix that the only combination of the eigenvectors that sums up to 0 at the boundary to satisfy the boundary condition is the trivial combination that the weights of all the eigenvectors are 0. 
For instance, returning to the example of Eq. \eqref{hamSSH1} and its eigenvectors Eq. \eqref{e0egv}, if $|c_{21}/d_{21}| < |d_{12}/c_{12}|$ then the corresponding square matrix formed by concatenating the first two eigenvectors is $\begin{pmatrix} 1 & 1 \\ 0 & 0 \end{pmatrix}$, which has a rank less than 2 and a linear superposition of these two eigenvectors that sums to zero can be found without the weights both equal to 0. In contrast, if the condition is not satisfied, the resulting matrix would be $\mathbf{I}_2$, which has a rank of 2. In this case, the system does not host nearly-zero energy edge modes because there is no non-trivial linear combination of the eigenvectors that sums to 0.

\bibliographystyle{apsrev4-2}
\bibliography{refSysSizeDep}

\begin{thebibliography}{72}%
\makeatletter
\providecommand \@ifxundefined [1]{%
 \@ifx{#1\undefined}
}%
\providecommand \@ifnum [1]{%
 \ifnum #1\expandafter \@firstoftwo
 \else \expandafter \@secondoftwo
 \fi
}%
\providecommand \@ifx [1]{%
 \ifx #1\expandafter \@firstoftwo
 \else \expandafter \@secondoftwo
 \fi
}%
\providecommand \natexlab [1]{#1}%
\providecommand \enquote  [1]{``#1''}%
\providecommand \bibnamefont  [1]{#1}%
\providecommand \bibfnamefont [1]{#1}%
\providecommand \citenamefont [1]{#1}%
\providecommand \href@noop [0]{\@secondoftwo}%
\providecommand \href [0]{\begingroup \@sanitize@url \@href}%
\providecommand \@href[1]{\@@startlink{#1}\@@href}%
\providecommand \@@href[1]{\endgroup#1\@@endlink}%
\providecommand \@sanitize@url [0]{\catcode `\\12\catcode `\$12\catcode
  `\&12\catcode `\#12\catcode `\^12\catcode `\_12\catcode `\%12\relax}%
\providecommand \@@startlink[1]{}%
\providecommand \@@endlink[0]{}%
\providecommand \url  [0]{\begingroup\@sanitize@url \@url }%
\providecommand \@url [1]{\endgroup\@href {#1}{\urlprefix }}%
\providecommand \urlprefix  [0]{URL }%
\providecommand \Eprint [0]{\href }%
\providecommand \doibase [0]{https://doi.org/}%
\providecommand \selectlanguage [0]{\@gobble}%
\providecommand \bibinfo  [0]{\@secondoftwo}%
\providecommand \bibfield  [0]{\@secondoftwo}%
\providecommand \translation [1]{[#1]}%
\providecommand \BibitemOpen [0]{}%
\providecommand \bibitemStop [0]{}%
\providecommand \bibitemNoStop [0]{.\EOS\space}%
\providecommand \EOS [0]{\spacefactor3000\relax}%
\providecommand \BibitemShut  [1]{\csname bibitem#1\endcsname}%
\let\auto@bib@innerbib\@empty
\bibitem [{\citenamefont {Hafezi}\ \emph {et~al.}(2013)\citenamefont {Hafezi},
  \citenamefont {Mittal}, \citenamefont {Fan}, \citenamefont {Migdall},\ and\
  \citenamefont {Taylor}}]{hafezi2013imaging}%
  \BibitemOpen
  \bibfield  {author} {\bibinfo {author} {\bibfnamefont {M.}~\bibnamefont
  {Hafezi}}, \bibinfo {author} {\bibfnamefont {S.}~\bibnamefont {Mittal}},
  \bibinfo {author} {\bibfnamefont {J.}~\bibnamefont {Fan}}, \bibinfo {author}
  {\bibfnamefont {A.}~\bibnamefont {Migdall}},\ and\ \bibinfo {author}
  {\bibfnamefont {J.}~\bibnamefont {Taylor}},\ }\href@noop {} {\bibfield
  {journal} {\bibinfo  {journal} {Nat. Photon.}\ }\textbf {\bibinfo {volume}
  {7}},\ \bibinfo {pages} {1001} (\bibinfo {year} {2013})}\BibitemShut
  {NoStop}%
\bibitem [{\citenamefont {Obana}\ \emph {et~al.}(2019)\citenamefont {Obana},
  \citenamefont {Liu},\ and\ \citenamefont
  {Wakabayashi}}]{obana2019topological}%
  \BibitemOpen
  \bibfield  {author} {\bibinfo {author} {\bibfnamefont {D.}~\bibnamefont
  {Obana}}, \bibinfo {author} {\bibfnamefont {F.}~\bibnamefont {Liu}},\ and\
  \bibinfo {author} {\bibfnamefont {K.}~\bibnamefont {Wakabayashi}},\
  }\href@noop {} {\bibfield  {journal} {\bibinfo  {journal} {Phys. Rev. B}\
  }\textbf {\bibinfo {volume} {100}},\ \bibinfo {pages} {075437} (\bibinfo
  {year} {2019})}\BibitemShut {NoStop}%
\bibitem [{\citenamefont {St-Jean}\ \emph {et~al.}(2017)\citenamefont
  {St-Jean}, \citenamefont {Goblot}, \citenamefont {Galopin}, \citenamefont
  {Lema{\^\i}tre}, \citenamefont {Ozawa}, \citenamefont {Le~Gratiet},
  \citenamefont {Sagnes}, \citenamefont {Bloch},\ and\ \citenamefont
  {Amo}}]{st2017lasing}%
  \BibitemOpen
  \bibfield  {author} {\bibinfo {author} {\bibfnamefont {P.}~\bibnamefont
  {St-Jean}}, \bibinfo {author} {\bibfnamefont {V.}~\bibnamefont {Goblot}},
  \bibinfo {author} {\bibfnamefont {E.}~\bibnamefont {Galopin}}, \bibinfo
  {author} {\bibfnamefont {A.}~\bibnamefont {Lema{\^\i}tre}}, \bibinfo {author}
  {\bibfnamefont {T.}~\bibnamefont {Ozawa}}, \bibinfo {author} {\bibfnamefont
  {L.}~\bibnamefont {Le~Gratiet}}, \bibinfo {author} {\bibfnamefont
  {I.}~\bibnamefont {Sagnes}}, \bibinfo {author} {\bibfnamefont
  {J.}~\bibnamefont {Bloch}},\ and\ \bibinfo {author} {\bibfnamefont
  {A.}~\bibnamefont {Amo}},\ }\href@noop {} {\bibfield  {journal} {\bibinfo
  {journal} {Nat. Photon.}\ }\textbf {\bibinfo {volume} {11}},\ \bibinfo
  {pages} {651} (\bibinfo {year} {2017})}\BibitemShut {NoStop}%
\bibitem [{\citenamefont {Rafi-Ul-Islam}\ \emph
  {et~al.}(2021{\natexlab{a}})\citenamefont {Rafi-Ul-Islam}, \citenamefont
  {Siu},\ and\ \citenamefont {Jalil}}]{rafi2021topological}%
  \BibitemOpen
  \bibfield  {author} {\bibinfo {author} {\bibfnamefont {S.}~\bibnamefont
  {Rafi-Ul-Islam}}, \bibinfo {author} {\bibfnamefont {Z.~B.}\ \bibnamefont
  {Siu}},\ and\ \bibinfo {author} {\bibfnamefont {M.~B.}\ \bibnamefont
  {Jalil}},\ }\href@noop {} {\bibfield  {journal} {\bibinfo  {journal} {Phys.
  Rev. B}\ }\textbf {\bibinfo {volume} {103}},\ \bibinfo {pages} {035420}
  (\bibinfo {year} {2021}{\natexlab{a}})}\BibitemShut {NoStop}%
\bibitem [{\citenamefont {Gao}\ \emph {et~al.}(2020{\natexlab{a}})\citenamefont
  {Gao}, \citenamefont {Xue}, \citenamefont {Wang}, \citenamefont {Gu},
  \citenamefont {Liu}, \citenamefont {Zhu},\ and\ \citenamefont
  {Zhang}}]{gao2020observation}%
  \BibitemOpen
  \bibfield  {author} {\bibinfo {author} {\bibfnamefont {H.}~\bibnamefont
  {Gao}}, \bibinfo {author} {\bibfnamefont {H.}~\bibnamefont {Xue}}, \bibinfo
  {author} {\bibfnamefont {Q.}~\bibnamefont {Wang}}, \bibinfo {author}
  {\bibfnamefont {Z.}~\bibnamefont {Gu}}, \bibinfo {author} {\bibfnamefont
  {T.}~\bibnamefont {Liu}}, \bibinfo {author} {\bibfnamefont {J.}~\bibnamefont
  {Zhu}},\ and\ \bibinfo {author} {\bibfnamefont {B.}~\bibnamefont {Zhang}},\
  }\href@noop {} {\bibfield  {journal} {\bibinfo  {journal} {Phys. Rev. B}\
  }\textbf {\bibinfo {volume} {101}},\ \bibinfo {pages} {180303} (\bibinfo
  {year} {2020}{\natexlab{a}})}\BibitemShut {NoStop}%
\bibitem [{\citenamefont {Ganeshan}\ \emph {et~al.}(2013)\citenamefont
  {Ganeshan}, \citenamefont {Sun},\ and\ \citenamefont
  {Sarma}}]{ganeshan2013topological}%
  \BibitemOpen
  \bibfield  {author} {\bibinfo {author} {\bibfnamefont {S.}~\bibnamefont
  {Ganeshan}}, \bibinfo {author} {\bibfnamefont {K.}~\bibnamefont {Sun}},\ and\
  \bibinfo {author} {\bibfnamefont {S.~D.}\ \bibnamefont {Sarma}},\ }\href@noop
  {} {\bibfield  {journal} {\bibinfo  {journal} {Phys. Rev. Lett.}\ }\textbf
  {\bibinfo {volume} {110}},\ \bibinfo {pages} {180403} (\bibinfo {year}
  {2013})}\BibitemShut {NoStop}%
\bibitem [{\citenamefont {Kempkes}\ \emph {et~al.}(2019)\citenamefont
  {Kempkes}, \citenamefont {Slot}, \citenamefont {van Den~Broeke},
  \citenamefont {Capiod}, \citenamefont {Benalcazar}, \citenamefont
  {Vanmaekelbergh}, \citenamefont {Bercioux}, \citenamefont {Swart},\ and\
  \citenamefont {Smith}}]{kempkes2019robust}%
  \BibitemOpen
  \bibfield  {author} {\bibinfo {author} {\bibfnamefont {S.}~\bibnamefont
  {Kempkes}}, \bibinfo {author} {\bibfnamefont {M.}~\bibnamefont {Slot}},
  \bibinfo {author} {\bibfnamefont {J.}~\bibnamefont {van Den~Broeke}},
  \bibinfo {author} {\bibfnamefont {P.}~\bibnamefont {Capiod}}, \bibinfo
  {author} {\bibfnamefont {W.}~\bibnamefont {Benalcazar}}, \bibinfo {author}
  {\bibfnamefont {D.}~\bibnamefont {Vanmaekelbergh}}, \bibinfo {author}
  {\bibfnamefont {D.}~\bibnamefont {Bercioux}}, \bibinfo {author}
  {\bibfnamefont {I.}~\bibnamefont {Swart}},\ and\ \bibinfo {author}
  {\bibfnamefont {C.~M.}\ \bibnamefont {Smith}},\ }\href@noop {} {\bibfield
  {journal} {\bibinfo  {journal} {Nat. Mater.}\ }\textbf {\bibinfo {volume}
  {18}},\ \bibinfo {pages} {1292} (\bibinfo {year} {2019})}\BibitemShut
  {NoStop}%
\bibitem [{\citenamefont {Ryu}\ and\ \citenamefont
  {Hatsugai}(2002)}]{ryu2002topological}%
  \BibitemOpen
  \bibfield  {author} {\bibinfo {author} {\bibfnamefont {S.}~\bibnamefont
  {Ryu}}\ and\ \bibinfo {author} {\bibfnamefont {Y.}~\bibnamefont {Hatsugai}},\
  }\href@noop {} {\bibfield  {journal} {\bibinfo  {journal} {Phys. Rev. Lett.}\
  }\textbf {\bibinfo {volume} {89}},\ \bibinfo {pages} {077002} (\bibinfo
  {year} {2002})}\BibitemShut {NoStop}%
\bibitem [{\citenamefont {Imhof}\ \emph {et~al.}(2018)\citenamefont {Imhof},
  \citenamefont {Berger}, \citenamefont {Bayer}, \citenamefont {Brehm},
  \citenamefont {Molenkamp}, \citenamefont {Kiessling}, \citenamefont
  {Schindler}, \citenamefont {Lee}, \citenamefont {Greiter}, \citenamefont
  {Neupert} \emph {et~al.}}]{imhof2018topolectrical}%
  \BibitemOpen
  \bibfield  {author} {\bibinfo {author} {\bibfnamefont {S.}~\bibnamefont
  {Imhof}}, \bibinfo {author} {\bibfnamefont {C.}~\bibnamefont {Berger}},
  \bibinfo {author} {\bibfnamefont {F.}~\bibnamefont {Bayer}}, \bibinfo
  {author} {\bibfnamefont {J.}~\bibnamefont {Brehm}}, \bibinfo {author}
  {\bibfnamefont {L.~W.}\ \bibnamefont {Molenkamp}}, \bibinfo {author}
  {\bibfnamefont {T.}~\bibnamefont {Kiessling}}, \bibinfo {author}
  {\bibfnamefont {F.}~\bibnamefont {Schindler}}, \bibinfo {author}
  {\bibfnamefont {C.~H.}\ \bibnamefont {Lee}}, \bibinfo {author} {\bibfnamefont
  {M.}~\bibnamefont {Greiter}}, \bibinfo {author} {\bibfnamefont
  {T.}~\bibnamefont {Neupert}}, \emph {et~al.},\ }\href@noop {} {\bibfield
  {journal} {\bibinfo  {journal} {Nat. Phys.}\ }\textbf {\bibinfo {volume}
  {14}},\ \bibinfo {pages} {925} (\bibinfo {year} {2018})}\BibitemShut
  {NoStop}%
\bibitem [{\citenamefont {Sahin}\ \emph
  {et~al.}(2022{\natexlab{a}})\citenamefont {Sahin}, \citenamefont
  {Rafi-Ul-Islam}, \citenamefont {Siu}, \citenamefont {Lee},\ and\
  \citenamefont {Jalil}}]{sahin2022interfacial}%
  \BibitemOpen
  \bibfield  {author} {\bibinfo {author} {\bibfnamefont {H.}~\bibnamefont
  {Sahin}}, \bibinfo {author} {\bibfnamefont {S.}~\bibnamefont
  {Rafi-Ul-Islam}}, \bibinfo {author} {\bibfnamefont {Z.~B.}\ \bibnamefont
  {Siu}}, \bibinfo {author} {\bibfnamefont {C.~H.}\ \bibnamefont {Lee}},\ and\
  \bibinfo {author} {\bibfnamefont {M.}~\bibnamefont {Jalil}},\ }\href@noop {}
  {\bibfield  {journal} {\bibinfo  {journal} {Bull. Am. Phys. Soc.}\ }
  (\bibinfo {year} {2022}{\natexlab{a}})}\BibitemShut {NoStop}%
\bibitem [{\citenamefont {Banerjee}\ \emph {et~al.}(2020)\citenamefont
  {Banerjee}, \citenamefont {Mandal},\ and\ \citenamefont
  {Liew}}]{banerjee2020coupling}%
  \BibitemOpen
  \bibfield  {author} {\bibinfo {author} {\bibfnamefont {R.}~\bibnamefont
  {Banerjee}}, \bibinfo {author} {\bibfnamefont {S.}~\bibnamefont {Mandal}},\
  and\ \bibinfo {author} {\bibfnamefont {T.}~\bibnamefont {Liew}},\ }\href@noop
  {} {\bibfield  {journal} {\bibinfo  {journal} {Phys. Rev. Lett.}\ }\textbf
  {\bibinfo {volume} {124}},\ \bibinfo {pages} {063901} (\bibinfo {year}
  {2020})}\BibitemShut {NoStop}%
\bibitem [{\citenamefont {Wu}\ \emph {et~al.}(2021)\citenamefont {Wu},
  \citenamefont {Meng}, \citenamefont {Hao}, \citenamefont {Zhang},
  \citenamefont {Li},\ and\ \citenamefont {Zhang}}]{wu2021topological}%
  \BibitemOpen
  \bibfield  {author} {\bibinfo {author} {\bibfnamefont {X.}~\bibnamefont
  {Wu}}, \bibinfo {author} {\bibfnamefont {Y.}~\bibnamefont {Meng}}, \bibinfo
  {author} {\bibfnamefont {Y.}~\bibnamefont {Hao}}, \bibinfo {author}
  {\bibfnamefont {R.-Y.}\ \bibnamefont {Zhang}}, \bibinfo {author}
  {\bibfnamefont {J.}~\bibnamefont {Li}},\ and\ \bibinfo {author}
  {\bibfnamefont {X.}~\bibnamefont {Zhang}},\ }\href@noop {} {\bibfield
  {journal} {\bibinfo  {journal} {Phys. Rev. Lett.}\ }\textbf {\bibinfo
  {volume} {126}},\ \bibinfo {pages} {226802} (\bibinfo {year}
  {2021})}\BibitemShut {NoStop}%
\bibitem [{\citenamefont {Kunst}\ \emph
  {et~al.}(2018{\natexlab{a}})\citenamefont {Kunst}, \citenamefont {van
  Miert},\ and\ \citenamefont {Bergholtz}}]{kunst2018lattice}%
  \BibitemOpen
  \bibfield  {author} {\bibinfo {author} {\bibfnamefont {F.~K.}\ \bibnamefont
  {Kunst}}, \bibinfo {author} {\bibfnamefont {G.}~\bibnamefont {van Miert}},\
  and\ \bibinfo {author} {\bibfnamefont {E.~J.}\ \bibnamefont {Bergholtz}},\
  }\href@noop {} {\bibfield  {journal} {\bibinfo  {journal} {Phys. Rev. B}\
  }\textbf {\bibinfo {volume} {97}},\ \bibinfo {pages} {241405} (\bibinfo
  {year} {2018}{\natexlab{a}})}\BibitemShut {NoStop}%
\bibitem [{\citenamefont {Benalcazar}\ \emph {et~al.}(2017)\citenamefont
  {Benalcazar}, \citenamefont {Bernevig},\ and\ \citenamefont
  {Hughes}}]{benalcazar2017electric}%
  \BibitemOpen
  \bibfield  {author} {\bibinfo {author} {\bibfnamefont {W.~A.}\ \bibnamefont
  {Benalcazar}}, \bibinfo {author} {\bibfnamefont {B.~A.}\ \bibnamefont
  {Bernevig}},\ and\ \bibinfo {author} {\bibfnamefont {T.~L.}\ \bibnamefont
  {Hughes}},\ }\href@noop {} {\bibfield  {journal} {\bibinfo  {journal} {Phys.
  Rev. B}\ }\textbf {\bibinfo {volume} {96}},\ \bibinfo {pages} {245115}
  (\bibinfo {year} {2017})}\BibitemShut {NoStop}%
\bibitem [{\citenamefont {Ghorashi}\ \emph {et~al.}(2019)\citenamefont
  {Ghorashi}, \citenamefont {Hu}, \citenamefont {Hughes},\ and\ \citenamefont
  {Rossi}}]{ghorashi2019second}%
  \BibitemOpen
  \bibfield  {author} {\bibinfo {author} {\bibfnamefont {S.~A.~A.}\
  \bibnamefont {Ghorashi}}, \bibinfo {author} {\bibfnamefont {X.}~\bibnamefont
  {Hu}}, \bibinfo {author} {\bibfnamefont {T.~L.}\ \bibnamefont {Hughes}},\
  and\ \bibinfo {author} {\bibfnamefont {E.}~\bibnamefont {Rossi}},\
  }\href@noop {} {\bibfield  {journal} {\bibinfo  {journal} {Phys. Rev. B}\
  }\textbf {\bibinfo {volume} {100}},\ \bibinfo {pages} {020509} (\bibinfo
  {year} {2019})}\BibitemShut {NoStop}%
\bibitem [{\citenamefont {Jin}\ \emph {et~al.}(2018)\citenamefont {Jin},
  \citenamefont {Torrent},\ and\ \citenamefont
  {Djafari-Rouhani}}]{jin2018robustness}%
  \BibitemOpen
  \bibfield  {author} {\bibinfo {author} {\bibfnamefont {Y.}~\bibnamefont
  {Jin}}, \bibinfo {author} {\bibfnamefont {D.}~\bibnamefont {Torrent}},\ and\
  \bibinfo {author} {\bibfnamefont {B.}~\bibnamefont {Djafari-Rouhani}},\
  }\href@noop {} {\bibfield  {journal} {\bibinfo  {journal} {Phys. Rev. B}\
  }\textbf {\bibinfo {volume} {98}},\ \bibinfo {pages} {054307} (\bibinfo
  {year} {2018})}\BibitemShut {NoStop}%
\bibitem [{\citenamefont {Fujita}\ \emph {et~al.}(2011)\citenamefont {Fujita},
  \citenamefont {Jalil}, \citenamefont {Tan},\ and\ \citenamefont
  {Murakami}}]{fujita2011gauge}%
  \BibitemOpen
  \bibfield  {author} {\bibinfo {author} {\bibfnamefont {T.}~\bibnamefont
  {Fujita}}, \bibinfo {author} {\bibfnamefont {M.}~\bibnamefont {Jalil}},
  \bibinfo {author} {\bibfnamefont {S.}~\bibnamefont {Tan}},\ and\ \bibinfo
  {author} {\bibfnamefont {S.}~\bibnamefont {Murakami}},\ }\href@noop {}
  {\bibfield  {journal} {\bibinfo  {journal} {J. Appl. Phys.}\ }\textbf
  {\bibinfo {volume} {110}},\ \bibinfo {pages} {17} (\bibinfo {year}
  {2011})}\BibitemShut {NoStop}%
\bibitem [{\citenamefont {Tan}\ \emph {et~al.}(2020)\citenamefont {Tan},
  \citenamefont {Chen}, \citenamefont {Ho}, \citenamefont {Huang},
  \citenamefont {Jalil}, \citenamefont {Chang},\ and\ \citenamefont
  {Murakami}}]{tan2020yang}%
  \BibitemOpen
  \bibfield  {author} {\bibinfo {author} {\bibfnamefont {S.~G.}\ \bibnamefont
  {Tan}}, \bibinfo {author} {\bibfnamefont {S.-H.}\ \bibnamefont {Chen}},
  \bibinfo {author} {\bibfnamefont {C.~S.}\ \bibnamefont {Ho}}, \bibinfo
  {author} {\bibfnamefont {C.-C.}\ \bibnamefont {Huang}}, \bibinfo {author}
  {\bibfnamefont {M.~B.}\ \bibnamefont {Jalil}}, \bibinfo {author}
  {\bibfnamefont {C.~R.}\ \bibnamefont {Chang}},\ and\ \bibinfo {author}
  {\bibfnamefont {S.}~\bibnamefont {Murakami}},\ }\href@noop {} {\bibfield
  {journal} {\bibinfo  {journal} {Phys. Rep.}\ }\textbf {\bibinfo {volume}
  {882}},\ \bibinfo {pages} {1} (\bibinfo {year} {2020})}\BibitemShut {NoStop}%
\bibitem [{\citenamefont {Wu}\ \emph {et~al.}(2017)\citenamefont {Wu},
  \citenamefont {Li}, \citenamefont {Hu}, \citenamefont {Ao}, \citenamefont
  {Zhao},\ and\ \citenamefont {Gong}}]{wu2017applications}%
  \BibitemOpen
  \bibfield  {author} {\bibinfo {author} {\bibfnamefont {Y.}~\bibnamefont
  {Wu}}, \bibinfo {author} {\bibfnamefont {C.}~\bibnamefont {Li}}, \bibinfo
  {author} {\bibfnamefont {X.}~\bibnamefont {Hu}}, \bibinfo {author}
  {\bibfnamefont {Y.}~\bibnamefont {Ao}}, \bibinfo {author} {\bibfnamefont
  {Y.}~\bibnamefont {Zhao}},\ and\ \bibinfo {author} {\bibfnamefont
  {Q.}~\bibnamefont {Gong}},\ }\href@noop {} {\bibfield  {journal} {\bibinfo
  {journal} {Adv. Opt. Mater.}\ }\textbf {\bibinfo {volume} {5}},\ \bibinfo
  {pages} {1700357} (\bibinfo {year} {2017})}\BibitemShut {NoStop}%
\bibitem [{\citenamefont {Vobornik}\ \emph {et~al.}(2011)\citenamefont
  {Vobornik}, \citenamefont {Manju}, \citenamefont {Fujii}, \citenamefont
  {Borgatti}, \citenamefont {Torelli}, \citenamefont {Krizmancic},
  \citenamefont {Hor}, \citenamefont {Cava},\ and\ \citenamefont
  {Panaccione}}]{vobornik2011magnetic}%
  \BibitemOpen
  \bibfield  {author} {\bibinfo {author} {\bibfnamefont {I.}~\bibnamefont
  {Vobornik}}, \bibinfo {author} {\bibfnamefont {U.}~\bibnamefont {Manju}},
  \bibinfo {author} {\bibfnamefont {J.}~\bibnamefont {Fujii}}, \bibinfo
  {author} {\bibfnamefont {F.}~\bibnamefont {Borgatti}}, \bibinfo {author}
  {\bibfnamefont {P.}~\bibnamefont {Torelli}}, \bibinfo {author} {\bibfnamefont
  {D.}~\bibnamefont {Krizmancic}}, \bibinfo {author} {\bibfnamefont {Y.~S.}\
  \bibnamefont {Hor}}, \bibinfo {author} {\bibfnamefont {R.~J.}\ \bibnamefont
  {Cava}},\ and\ \bibinfo {author} {\bibfnamefont {G.}~\bibnamefont
  {Panaccione}},\ }\href@noop {} {\bibfield  {journal} {\bibinfo  {journal}
  {Nano Lett.}\ }\textbf {\bibinfo {volume} {11}},\ \bibinfo {pages} {4079}
  (\bibinfo {year} {2011})}\BibitemShut {NoStop}%
\bibitem [{\citenamefont {Kesselring}\ \emph {et~al.}(2018)\citenamefont
  {Kesselring}, \citenamefont {Pastawski}, \citenamefont {Eisert},\ and\
  \citenamefont {Brown}}]{kesselring2018boundaries}%
  \BibitemOpen
  \bibfield  {author} {\bibinfo {author} {\bibfnamefont {M.~S.}\ \bibnamefont
  {Kesselring}}, \bibinfo {author} {\bibfnamefont {F.}~\bibnamefont
  {Pastawski}}, \bibinfo {author} {\bibfnamefont {J.}~\bibnamefont {Eisert}},\
  and\ \bibinfo {author} {\bibfnamefont {B.~J.}\ \bibnamefont {Brown}},\
  }\href@noop {} {\bibfield  {journal} {\bibinfo  {journal} {Quantum}\ }\textbf
  {\bibinfo {volume} {2}},\ \bibinfo {pages} {101} (\bibinfo {year}
  {2018})}\BibitemShut {NoStop}%
\bibitem [{\citenamefont {Helbig}\ \emph {et~al.}(2020)\citenamefont {Helbig},
  \citenamefont {Hofmann}, \citenamefont {Imhof}, \citenamefont {Abdelghany},
  \citenamefont {Kiessling}, \citenamefont {Molenkamp}, \citenamefont {Lee},
  \citenamefont {Szameit}, \citenamefont {Greiter},\ and\ \citenamefont
  {Thomale}}]{helbig2020generalized}%
  \BibitemOpen
  \bibfield  {author} {\bibinfo {author} {\bibfnamefont {T.}~\bibnamefont
  {Helbig}}, \bibinfo {author} {\bibfnamefont {T.}~\bibnamefont {Hofmann}},
  \bibinfo {author} {\bibfnamefont {S.}~\bibnamefont {Imhof}}, \bibinfo
  {author} {\bibfnamefont {M.}~\bibnamefont {Abdelghany}}, \bibinfo {author}
  {\bibfnamefont {T.}~\bibnamefont {Kiessling}}, \bibinfo {author}
  {\bibfnamefont {L.}~\bibnamefont {Molenkamp}}, \bibinfo {author}
  {\bibfnamefont {C.}~\bibnamefont {Lee}}, \bibinfo {author} {\bibfnamefont
  {A.}~\bibnamefont {Szameit}}, \bibinfo {author} {\bibfnamefont
  {M.}~\bibnamefont {Greiter}},\ and\ \bibinfo {author} {\bibfnamefont
  {R.}~\bibnamefont {Thomale}},\ }\href@noop {} {\bibfield  {journal} {\bibinfo
   {journal} {Nat. Phys.}\ }\textbf {\bibinfo {volume} {16}},\ \bibinfo {pages}
  {747} (\bibinfo {year} {2020})}\BibitemShut {NoStop}%
\bibitem [{\citenamefont {Qiu}\ \emph {et~al.}(2019)\citenamefont {Qiu},
  \citenamefont {Bai}, \citenamefont {Zhang}, \citenamefont {Xin},
  \citenamefont {Zou}, \citenamefont {Zhang}, \citenamefont {Jin},
  \citenamefont {An},\ and\ \citenamefont {Zhang}}]{qiu2019unidirectional}%
  \BibitemOpen
  \bibfield  {author} {\bibinfo {author} {\bibfnamefont {D.-X.}\ \bibnamefont
  {Qiu}}, \bibinfo {author} {\bibfnamefont {R.}~\bibnamefont {Bai}}, \bibinfo
  {author} {\bibfnamefont {C.}~\bibnamefont {Zhang}}, \bibinfo {author}
  {\bibfnamefont {L.-F.}\ \bibnamefont {Xin}}, \bibinfo {author} {\bibfnamefont
  {X.-Y.}\ \bibnamefont {Zou}}, \bibinfo {author} {\bibfnamefont {Y.~Q.}\
  \bibnamefont {Zhang}}, \bibinfo {author} {\bibfnamefont {X.~R.}\ \bibnamefont
  {Jin}}, \bibinfo {author} {\bibfnamefont {C.}~\bibnamefont {An}},\ and\
  \bibinfo {author} {\bibfnamefont {S.}~\bibnamefont {Zhang}},\ }\href@noop {}
  {\bibfield  {journal} {\bibinfo  {journal} {Quant. Inf. Process.}\ }\textbf
  {\bibinfo {volume} {18}},\ \bibinfo {pages} {1} (\bibinfo {year}
  {2019})}\BibitemShut {NoStop}%
\bibitem [{\citenamefont {Santos}\ and\ \citenamefont
  {da~Silva}(2014)}]{santos2014non}%
  \BibitemOpen
  \bibfield  {author} {\bibinfo {author} {\bibfnamefont {R.~B.}\ \bibnamefont
  {Santos}}\ and\ \bibinfo {author} {\bibfnamefont {V.~R.}\ \bibnamefont
  {da~Silva}},\ }\href@noop {} {\bibfield  {journal} {\bibinfo  {journal} {Mod.
  Phys. Lett. B}\ }\textbf {\bibinfo {volume} {28}},\ \bibinfo {pages}
  {1450223} (\bibinfo {year} {2014})}\BibitemShut {NoStop}%
\bibitem [{\citenamefont {Rafi-Ul-Islam}\ \emph
  {et~al.}(2021{\natexlab{b}})\citenamefont {Rafi-Ul-Islam}, \citenamefont
  {Siu},\ and\ \citenamefont {Jalil}}]{rafi2021non}%
  \BibitemOpen
  \bibfield  {author} {\bibinfo {author} {\bibfnamefont {S.}~\bibnamefont
  {Rafi-Ul-Islam}}, \bibinfo {author} {\bibfnamefont {Z.~B.}\ \bibnamefont
  {Siu}},\ and\ \bibinfo {author} {\bibfnamefont {M.~B.}\ \bibnamefont
  {Jalil}},\ }\href@noop {} {\bibfield  {journal} {\bibinfo  {journal} {New J.
  Phys.}\ }\textbf {\bibinfo {volume} {23}},\ \bibinfo {pages} {033014}
  (\bibinfo {year} {2021}{\natexlab{b}})}\BibitemShut {NoStop}%
\bibitem [{\citenamefont {Li}\ and\ \citenamefont {Lee}(2022)}]{li2022non}%
  \BibitemOpen
  \bibfield  {author} {\bibinfo {author} {\bibfnamefont {L.}~\bibnamefont
  {Li}}\ and\ \bibinfo {author} {\bibfnamefont {C.~H.}\ \bibnamefont {Lee}},\
  }\href@noop {} {\bibfield  {journal} {\bibinfo  {journal} {Sci. Bull.}\ }
  (\bibinfo {year} {2022})}\BibitemShut {NoStop}%
\bibitem [{\citenamefont {Yang}\ \emph {et~al.}(2022)\citenamefont {Yang},
  \citenamefont {Tan}, \citenamefont {Tai}, \citenamefont {Koh}, \citenamefont
  {Li}, \citenamefont {Longhi},\ and\ \citenamefont {Lee}}]{yang2022designing}%
  \BibitemOpen
  \bibfield  {author} {\bibinfo {author} {\bibfnamefont {R.}~\bibnamefont
  {Yang}}, \bibinfo {author} {\bibfnamefont {J.~W.}\ \bibnamefont {Tan}},
  \bibinfo {author} {\bibfnamefont {T.}~\bibnamefont {Tai}}, \bibinfo {author}
  {\bibfnamefont {J.~M.}\ \bibnamefont {Koh}}, \bibinfo {author} {\bibfnamefont
  {L.}~\bibnamefont {Li}}, \bibinfo {author} {\bibfnamefont {S.}~\bibnamefont
  {Longhi}},\ and\ \bibinfo {author} {\bibfnamefont {C.~H.}\ \bibnamefont
  {Lee}},\ }\href@noop {} {\bibfield  {journal} {\bibinfo  {journal} {arXiv
  preprint arXiv:2201.04153}\ } (\bibinfo {year} {2022})}\BibitemShut {NoStop}%
\bibitem [{\citenamefont {Zeuner}\ \emph {et~al.}(2015)\citenamefont {Zeuner},
  \citenamefont {Rechtsman}, \citenamefont {Plotnik}, \citenamefont {Lumer},
  \citenamefont {Nolte}, \citenamefont {Rudner}, \citenamefont {Segev},\ and\
  \citenamefont {Szameit}}]{zeuner2015observation}%
  \BibitemOpen
  \bibfield  {author} {\bibinfo {author} {\bibfnamefont {J.~M.}\ \bibnamefont
  {Zeuner}}, \bibinfo {author} {\bibfnamefont {M.~C.}\ \bibnamefont
  {Rechtsman}}, \bibinfo {author} {\bibfnamefont {Y.}~\bibnamefont {Plotnik}},
  \bibinfo {author} {\bibfnamefont {Y.}~\bibnamefont {Lumer}}, \bibinfo
  {author} {\bibfnamefont {S.}~\bibnamefont {Nolte}}, \bibinfo {author}
  {\bibfnamefont {M.~S.}\ \bibnamefont {Rudner}}, \bibinfo {author}
  {\bibfnamefont {M.}~\bibnamefont {Segev}},\ and\ \bibinfo {author}
  {\bibfnamefont {A.}~\bibnamefont {Szameit}},\ }\href@noop {} {\bibfield
  {journal} {\bibinfo  {journal} {Phys. Rev. Lett.}\ }\textbf {\bibinfo
  {volume} {115}},\ \bibinfo {pages} {040402} (\bibinfo {year}
  {2015})}\BibitemShut {NoStop}%
\bibitem [{\citenamefont {Tai}\ and\ \citenamefont
  {Lee}(2022)}]{tai2022zoology}%
  \BibitemOpen
  \bibfield  {author} {\bibinfo {author} {\bibfnamefont {T.}~\bibnamefont
  {Tai}}\ and\ \bibinfo {author} {\bibfnamefont {C.~H.}\ \bibnamefont {Lee}},\
  }\href@noop {} {\bibfield  {journal} {\bibinfo  {journal} {arXiv preprint
  arXiv:2202.03462}\ } (\bibinfo {year} {2022})}\BibitemShut {NoStop}%
\bibitem [{\citenamefont {Li}\ \emph {et~al.}(2021{\natexlab{a}})\citenamefont
  {Li}, \citenamefont {Mu}, \citenamefont {Lee},\ and\ \citenamefont
  {Gong}}]{li2021quantized}%
  \BibitemOpen
  \bibfield  {author} {\bibinfo {author} {\bibfnamefont {L.}~\bibnamefont
  {Li}}, \bibinfo {author} {\bibfnamefont {S.}~\bibnamefont {Mu}}, \bibinfo
  {author} {\bibfnamefont {C.~H.}\ \bibnamefont {Lee}},\ and\ \bibinfo {author}
  {\bibfnamefont {J.}~\bibnamefont {Gong}},\ }\href@noop {} {\bibfield
  {journal} {\bibinfo  {journal} {Nat. Commun.}\ }\textbf {\bibinfo {volume}
  {12}},\ \bibinfo {pages} {1} (\bibinfo {year}
  {2021}{\natexlab{a}})}\BibitemShut {NoStop}%
\bibitem [{\citenamefont {Gu}\ \emph {et~al.}(2016)\citenamefont {Gu},
  \citenamefont {Lee}, \citenamefont {Wen}, \citenamefont {Cho}, \citenamefont
  {Ryu},\ and\ \citenamefont {Qi}}]{gu2016holographic}%
  \BibitemOpen
  \bibfield  {author} {\bibinfo {author} {\bibfnamefont {Y.}~\bibnamefont
  {Gu}}, \bibinfo {author} {\bibfnamefont {C.~H.}\ \bibnamefont {Lee}},
  \bibinfo {author} {\bibfnamefont {X.}~\bibnamefont {Wen}}, \bibinfo {author}
  {\bibfnamefont {G.~Y.}\ \bibnamefont {Cho}}, \bibinfo {author} {\bibfnamefont
  {S.}~\bibnamefont {Ryu}},\ and\ \bibinfo {author} {\bibfnamefont {X.-L.}\
  \bibnamefont {Qi}},\ }\href@noop {} {\bibfield  {journal} {\bibinfo
  {journal} {Phys. Rev. B}\ }\textbf {\bibinfo {volume} {94}},\ \bibinfo
  {pages} {125107} (\bibinfo {year} {2016})}\BibitemShut {NoStop}%
\bibitem [{\citenamefont {Leykam}\ \emph {et~al.}(2017)\citenamefont {Leykam},
  \citenamefont {Bliokh}, \citenamefont {Huang}, \citenamefont {Chong},\ and\
  \citenamefont {Nori}}]{leykam2017edge}%
  \BibitemOpen
  \bibfield  {author} {\bibinfo {author} {\bibfnamefont {D.}~\bibnamefont
  {Leykam}}, \bibinfo {author} {\bibfnamefont {K.~Y.}\ \bibnamefont {Bliokh}},
  \bibinfo {author} {\bibfnamefont {C.}~\bibnamefont {Huang}}, \bibinfo
  {author} {\bibfnamefont {Y.~D.}\ \bibnamefont {Chong}},\ and\ \bibinfo
  {author} {\bibfnamefont {F.}~\bibnamefont {Nori}},\ }\href@noop {} {\bibfield
   {journal} {\bibinfo  {journal} {Phys. Rev. Lett.}\ }\textbf {\bibinfo
  {volume} {118}},\ \bibinfo {pages} {040401} (\bibinfo {year}
  {2017})}\BibitemShut {NoStop}%
\bibitem [{\citenamefont {Gong}\ \emph {et~al.}(2018)\citenamefont {Gong},
  \citenamefont {Ashida}, \citenamefont {Kawabata}, \citenamefont {Takasan},
  \citenamefont {Higashikawa},\ and\ \citenamefont
  {Ueda}}]{gong2018topological}%
  \BibitemOpen
  \bibfield  {author} {\bibinfo {author} {\bibfnamefont {Z.}~\bibnamefont
  {Gong}}, \bibinfo {author} {\bibfnamefont {Y.}~\bibnamefont {Ashida}},
  \bibinfo {author} {\bibfnamefont {K.}~\bibnamefont {Kawabata}}, \bibinfo
  {author} {\bibfnamefont {K.}~\bibnamefont {Takasan}}, \bibinfo {author}
  {\bibfnamefont {S.}~\bibnamefont {Higashikawa}},\ and\ \bibinfo {author}
  {\bibfnamefont {M.}~\bibnamefont {Ueda}},\ }\href@noop {} {\bibfield
  {journal} {\bibinfo  {journal} {Phys. Rev. X}\ }\textbf {\bibinfo {volume}
  {8}},\ \bibinfo {pages} {031079} (\bibinfo {year} {2018})}\BibitemShut
  {NoStop}%
\bibitem [{\citenamefont {El-Ganainy}\ \emph {et~al.}(2018)\citenamefont
  {El-Ganainy}, \citenamefont {Makris}, \citenamefont {Khajavikhan},
  \citenamefont {Musslimani}, \citenamefont {Rotter},\ and\ \citenamefont
  {Christodoulides}}]{el2018non}%
  \BibitemOpen
  \bibfield  {author} {\bibinfo {author} {\bibfnamefont {R.}~\bibnamefont
  {El-Ganainy}}, \bibinfo {author} {\bibfnamefont {K.~G.}\ \bibnamefont
  {Makris}}, \bibinfo {author} {\bibfnamefont {M.}~\bibnamefont {Khajavikhan}},
  \bibinfo {author} {\bibfnamefont {Z.~H.}\ \bibnamefont {Musslimani}},
  \bibinfo {author} {\bibfnamefont {S.}~\bibnamefont {Rotter}},\ and\ \bibinfo
  {author} {\bibfnamefont {D.~N.}\ \bibnamefont {Christodoulides}},\
  }\href@noop {} {\bibfield  {journal} {\bibinfo  {journal} {Nat. Phys.}\
  }\textbf {\bibinfo {volume} {14}},\ \bibinfo {pages} {11} (\bibinfo {year}
  {2018})}\BibitemShut {NoStop}%
\bibitem [{\citenamefont {Yokomizo}\ and\ \citenamefont
  {Murakami}(2019)}]{yokomizo2019non}%
  \BibitemOpen
  \bibfield  {author} {\bibinfo {author} {\bibfnamefont {K.}~\bibnamefont
  {Yokomizo}}\ and\ \bibinfo {author} {\bibfnamefont {S.}~\bibnamefont
  {Murakami}},\ }\href@noop {} {\bibfield  {journal} {\bibinfo  {journal}
  {Phys. Rev. Lett.}\ }\textbf {\bibinfo {volume} {123}},\ \bibinfo {pages}
  {066404} (\bibinfo {year} {2019})}\BibitemShut {NoStop}%
\bibitem [{\citenamefont {Rafi-Ul-Islam}\ \emph
  {et~al.}(2020{\natexlab{a}})\citenamefont {Rafi-Ul-Islam}, \citenamefont
  {Siu},\ and\ \citenamefont {Jalil}}]{rafi2020topoelectrical}%
  \BibitemOpen
  \bibfield  {author} {\bibinfo {author} {\bibfnamefont {S.}~\bibnamefont
  {Rafi-Ul-Islam}}, \bibinfo {author} {\bibfnamefont {Z.~B.}\ \bibnamefont
  {Siu}},\ and\ \bibinfo {author} {\bibfnamefont {M.~B.}\ \bibnamefont
  {Jalil}},\ }\href@noop {} {\bibfield  {journal} {\bibinfo  {journal} {Commun.
  Phys.}\ }\textbf {\bibinfo {volume} {3}},\ \bibinfo {pages} {1} (\bibinfo
  {year} {2020}{\natexlab{a}})}\BibitemShut {NoStop}%
\bibitem [{\citenamefont {Zhang}\ \emph {et~al.}(2022)\citenamefont {Zhang},
  \citenamefont {Zhang}, \citenamefont {Sahin}, \citenamefont {Siu},
  \citenamefont {Rafi-Ul-Islam}, \citenamefont {Kong}, \citenamefont {Jalil},
  \citenamefont {Thomale},\ and\ \citenamefont {Lee}}]{zhang2022anomalous}%
  \BibitemOpen
  \bibfield  {author} {\bibinfo {author} {\bibfnamefont {X.}~\bibnamefont
  {Zhang}}, \bibinfo {author} {\bibfnamefont {B.}~\bibnamefont {Zhang}},
  \bibinfo {author} {\bibfnamefont {H.}~\bibnamefont {Sahin}}, \bibinfo
  {author} {\bibfnamefont {Z.~B.}\ \bibnamefont {Siu}}, \bibinfo {author}
  {\bibfnamefont {S.}~\bibnamefont {Rafi-Ul-Islam}}, \bibinfo {author}
  {\bibfnamefont {J.~F.}\ \bibnamefont {Kong}}, \bibinfo {author}
  {\bibfnamefont {M.}~\bibnamefont {Jalil}}, \bibinfo {author} {\bibfnamefont
  {R.}~\bibnamefont {Thomale}},\ and\ \bibinfo {author} {\bibfnamefont {C.~H.}\
  \bibnamefont {Lee}},\ }\href@noop {} {\bibfield  {journal} {\bibinfo
  {journal} {arXiv preprint arXiv:2204.05329}\ } (\bibinfo {year}
  {2022})}\BibitemShut {NoStop}%
\bibitem [{\citenamefont {Rafi-Ul-Islam}\ \emph
  {et~al.}(2021{\natexlab{c}})\citenamefont {Rafi-Ul-Islam}, \citenamefont
  {Siu}, \citenamefont {Sahin}, \citenamefont {Lee},\ and\ \citenamefont
  {Jalil}}]{rafi2021unconventional}%
  \BibitemOpen
  \bibfield  {author} {\bibinfo {author} {\bibfnamefont {S.}~\bibnamefont
  {Rafi-Ul-Islam}}, \bibinfo {author} {\bibfnamefont {Z.~B.}\ \bibnamefont
  {Siu}}, \bibinfo {author} {\bibfnamefont {H.}~\bibnamefont {Sahin}}, \bibinfo
  {author} {\bibfnamefont {C.~H.}\ \bibnamefont {Lee}},\ and\ \bibinfo {author}
  {\bibfnamefont {M.}~\bibnamefont {Jalil}},\ }\href@noop {} {\bibfield
  {journal} {\bibinfo  {journal} {ArXiv Preprint ArXiv:2108.01366}\ } (\bibinfo
  {year} {2021}{\natexlab{c}})}\BibitemShut {NoStop}%
\bibitem [{\citenamefont {Lenggenhager}\ \emph {et~al.}(2021)\citenamefont
  {Lenggenhager}, \citenamefont {Stegmaier}, \citenamefont {Upreti},
  \citenamefont {Hofmann}, \citenamefont {Helbig}, \citenamefont {Vollhardt},
  \citenamefont {Greiter}, \citenamefont {Lee}, \citenamefont {Imhof},
  \citenamefont {Brand} \emph {et~al.}}]{lenggenhager2021electric}%
  \BibitemOpen
  \bibfield  {author} {\bibinfo {author} {\bibfnamefont {P.~M.}\ \bibnamefont
  {Lenggenhager}}, \bibinfo {author} {\bibfnamefont {A.}~\bibnamefont
  {Stegmaier}}, \bibinfo {author} {\bibfnamefont {L.~K.}\ \bibnamefont
  {Upreti}}, \bibinfo {author} {\bibfnamefont {T.}~\bibnamefont {Hofmann}},
  \bibinfo {author} {\bibfnamefont {T.}~\bibnamefont {Helbig}}, \bibinfo
  {author} {\bibfnamefont {A.}~\bibnamefont {Vollhardt}}, \bibinfo {author}
  {\bibfnamefont {M.}~\bibnamefont {Greiter}}, \bibinfo {author} {\bibfnamefont
  {C.~H.}\ \bibnamefont {Lee}}, \bibinfo {author} {\bibfnamefont
  {S.}~\bibnamefont {Imhof}}, \bibinfo {author} {\bibfnamefont
  {H.}~\bibnamefont {Brand}}, \emph {et~al.},\ }\href@noop {} {\bibfield
  {journal} {\bibinfo  {journal} {arXiv preprint arXiv:2109.01148}\ } (\bibinfo
  {year} {2021})}\BibitemShut {NoStop}%
\bibitem [{\citenamefont {Rafi-Ul-Islam}\ \emph
  {et~al.}(2020{\natexlab{b}})\citenamefont {Rafi-Ul-Islam}, \citenamefont
  {Siu},\ and\ \citenamefont {Jalil}}]{rafi2020anti}%
  \BibitemOpen
  \bibfield  {author} {\bibinfo {author} {\bibfnamefont {S.}~\bibnamefont
  {Rafi-Ul-Islam}}, \bibinfo {author} {\bibfnamefont {Z.~B.}\ \bibnamefont
  {Siu}},\ and\ \bibinfo {author} {\bibfnamefont {M.~B.}\ \bibnamefont
  {Jalil}},\ }\href@noop {} {\bibfield  {journal} {\bibinfo  {journal} {Appl.
  Phys. Lett.}\ }\textbf {\bibinfo {volume} {116}},\ \bibinfo {pages} {111904}
  (\bibinfo {year} {2020}{\natexlab{b}})}\BibitemShut {NoStop}%
\bibitem [{\citenamefont {Stegmaier}\ \emph {et~al.}(2021)\citenamefont
  {Stegmaier}, \citenamefont {Imhof}, \citenamefont {Helbig}, \citenamefont
  {Hofmann}, \citenamefont {Lee}, \citenamefont {Kremer}, \citenamefont
  {Fritzsche}, \citenamefont {Feichtner}, \citenamefont {Klembt}, \citenamefont
  {H{\"o}fling} \emph {et~al.}}]{stegmaier2021topological}%
  \BibitemOpen
  \bibfield  {author} {\bibinfo {author} {\bibfnamefont {A.}~\bibnamefont
  {Stegmaier}}, \bibinfo {author} {\bibfnamefont {S.}~\bibnamefont {Imhof}},
  \bibinfo {author} {\bibfnamefont {T.}~\bibnamefont {Helbig}}, \bibinfo
  {author} {\bibfnamefont {T.}~\bibnamefont {Hofmann}}, \bibinfo {author}
  {\bibfnamefont {C.~H.}\ \bibnamefont {Lee}}, \bibinfo {author} {\bibfnamefont
  {M.}~\bibnamefont {Kremer}}, \bibinfo {author} {\bibfnamefont
  {A.}~\bibnamefont {Fritzsche}}, \bibinfo {author} {\bibfnamefont
  {T.}~\bibnamefont {Feichtner}}, \bibinfo {author} {\bibfnamefont
  {S.}~\bibnamefont {Klembt}}, \bibinfo {author} {\bibfnamefont
  {S.}~\bibnamefont {H{\"o}fling}}, \emph {et~al.},\ }\href@noop {} {\bibfield
  {journal} {\bibinfo  {journal} {Phys. Rev. Lett.}\ }\textbf {\bibinfo
  {volume} {126}},\ \bibinfo {pages} {215302} (\bibinfo {year}
  {2021})}\BibitemShut {NoStop}%
\bibitem [{\citenamefont {Rafi-Ul-Islam}\ \emph
  {et~al.}(2020{\natexlab{c}})\citenamefont {Rafi-Ul-Islam}, \citenamefont
  {Siu}, \citenamefont {Sun},\ and\ \citenamefont
  {Jalil}}]{rafi2020realization}%
  \BibitemOpen
  \bibfield  {author} {\bibinfo {author} {\bibfnamefont {S.}~\bibnamefont
  {Rafi-Ul-Islam}}, \bibinfo {author} {\bibfnamefont {Z.~B.}\ \bibnamefont
  {Siu}}, \bibinfo {author} {\bibfnamefont {C.}~\bibnamefont {Sun}},\ and\
  \bibinfo {author} {\bibfnamefont {M.~B.}\ \bibnamefont {Jalil}},\ }\href@noop
  {} {\bibfield  {journal} {\bibinfo  {journal} {New J. Phys.}\ }\textbf
  {\bibinfo {volume} {22}},\ \bibinfo {pages} {023025} (\bibinfo {year}
  {2020}{\natexlab{c}})}\BibitemShut {NoStop}%
\bibitem [{\citenamefont {Lee}\ \emph {et~al.}(2018)\citenamefont {Lee},
  \citenamefont {Imhof}, \citenamefont {Berger}, \citenamefont {Bayer},
  \citenamefont {Brehm}, \citenamefont {Molenkamp}, \citenamefont {Kiessling},\
  and\ \citenamefont {Thomale}}]{lee2018topolectrical}%
  \BibitemOpen
  \bibfield  {author} {\bibinfo {author} {\bibfnamefont {C.~H.}\ \bibnamefont
  {Lee}}, \bibinfo {author} {\bibfnamefont {S.}~\bibnamefont {Imhof}}, \bibinfo
  {author} {\bibfnamefont {C.}~\bibnamefont {Berger}}, \bibinfo {author}
  {\bibfnamefont {F.}~\bibnamefont {Bayer}}, \bibinfo {author} {\bibfnamefont
  {J.}~\bibnamefont {Brehm}}, \bibinfo {author} {\bibfnamefont {L.~W.}\
  \bibnamefont {Molenkamp}}, \bibinfo {author} {\bibfnamefont {T.}~\bibnamefont
  {Kiessling}},\ and\ \bibinfo {author} {\bibfnamefont {R.}~\bibnamefont
  {Thomale}},\ }\href@noop {} {\bibfield  {journal} {\bibinfo  {journal}
  {Commun. Phys.}\ }\textbf {\bibinfo {volume} {1}},\ \bibinfo {pages} {1}
  (\bibinfo {year} {2018})}\BibitemShut {NoStop}%
\bibitem [{\citenamefont {Hofmann}\ \emph {et~al.}(2020)\citenamefont
  {Hofmann}, \citenamefont {Helbig}, \citenamefont {Schindler}, \citenamefont
  {Salgo}, \citenamefont {Brzezi{\'n}ska}, \citenamefont {Greiter},
  \citenamefont {Kiessling}, \citenamefont {Wolf}, \citenamefont {Vollhardt},
  \citenamefont {Kaba{\v{s}}i} \emph {et~al.}}]{hofmann2020reciprocal}%
  \BibitemOpen
  \bibfield  {author} {\bibinfo {author} {\bibfnamefont {T.}~\bibnamefont
  {Hofmann}}, \bibinfo {author} {\bibfnamefont {T.}~\bibnamefont {Helbig}},
  \bibinfo {author} {\bibfnamefont {F.}~\bibnamefont {Schindler}}, \bibinfo
  {author} {\bibfnamefont {N.}~\bibnamefont {Salgo}}, \bibinfo {author}
  {\bibfnamefont {M.}~\bibnamefont {Brzezi{\'n}ska}}, \bibinfo {author}
  {\bibfnamefont {M.}~\bibnamefont {Greiter}}, \bibinfo {author} {\bibfnamefont
  {T.}~\bibnamefont {Kiessling}}, \bibinfo {author} {\bibfnamefont
  {D.}~\bibnamefont {Wolf}}, \bibinfo {author} {\bibfnamefont {A.}~\bibnamefont
  {Vollhardt}}, \bibinfo {author} {\bibfnamefont {A.}~\bibnamefont
  {Kaba{\v{s}}i}}, \emph {et~al.},\ }\href@noop {} {\bibfield  {journal}
  {\bibinfo  {journal} {Phys. Rev. Res.}\ }\textbf {\bibinfo {volume} {2}},\
  \bibinfo {pages} {023265} (\bibinfo {year} {2020})}\BibitemShut {NoStop}%
\bibitem [{\citenamefont {Kotwal}\ \emph {et~al.}(2019)\citenamefont {Kotwal},
  \citenamefont {Ronellenfitsch}, \citenamefont {Moseley}, \citenamefont
  {Stegmaier}, \citenamefont {Thomale},\ and\ \citenamefont
  {Dunkel}}]{kotwal2019active}%
  \BibitemOpen
  \bibfield  {author} {\bibinfo {author} {\bibfnamefont {T.}~\bibnamefont
  {Kotwal}}, \bibinfo {author} {\bibfnamefont {H.}~\bibnamefont
  {Ronellenfitsch}}, \bibinfo {author} {\bibfnamefont {F.}~\bibnamefont
  {Moseley}}, \bibinfo {author} {\bibfnamefont {A.}~\bibnamefont {Stegmaier}},
  \bibinfo {author} {\bibfnamefont {R.}~\bibnamefont {Thomale}},\ and\ \bibinfo
  {author} {\bibfnamefont {J.}~\bibnamefont {Dunkel}},\ }\href@noop {}
  {\bibfield  {journal} {\bibinfo  {journal} {ArXiv Preprint ArXiv:1903.10130}\
  } (\bibinfo {year} {2019})}\BibitemShut {NoStop}%
\bibitem [{\citenamefont {Zhu}\ \emph {et~al.}(2020)\citenamefont {Zhu},
  \citenamefont {Wang}, \citenamefont {Gupta}, \citenamefont {Zhang},
  \citenamefont {Xie}, \citenamefont {Lu},\ and\ \citenamefont
  {Chen}}]{zhu2020photonic}%
  \BibitemOpen
  \bibfield  {author} {\bibinfo {author} {\bibfnamefont {X.}~\bibnamefont
  {Zhu}}, \bibinfo {author} {\bibfnamefont {H.}~\bibnamefont {Wang}}, \bibinfo
  {author} {\bibfnamefont {S.~K.}\ \bibnamefont {Gupta}}, \bibinfo {author}
  {\bibfnamefont {H.}~\bibnamefont {Zhang}}, \bibinfo {author} {\bibfnamefont
  {B.}~\bibnamefont {Xie}}, \bibinfo {author} {\bibfnamefont {M.}~\bibnamefont
  {Lu}},\ and\ \bibinfo {author} {\bibfnamefont {Y.}~\bibnamefont {Chen}},\
  }\href@noop {} {\bibfield  {journal} {\bibinfo  {journal} {Phys. Rev. Res.}\
  }\textbf {\bibinfo {volume} {2}},\ \bibinfo {pages} {013280} (\bibinfo {year}
  {2020})}\BibitemShut {NoStop}%
\bibitem [{\citenamefont {Song}\ \emph {et~al.}(2020)\citenamefont {Song},
  \citenamefont {Liu}, \citenamefont {Zheng}, \citenamefont {Zhang},
  \citenamefont {Wang},\ and\ \citenamefont {Lu}}]{song2020two}%
  \BibitemOpen
  \bibfield  {author} {\bibinfo {author} {\bibfnamefont {Y.}~\bibnamefont
  {Song}}, \bibinfo {author} {\bibfnamefont {W.}~\bibnamefont {Liu}}, \bibinfo
  {author} {\bibfnamefont {L.}~\bibnamefont {Zheng}}, \bibinfo {author}
  {\bibfnamefont {Y.}~\bibnamefont {Zhang}}, \bibinfo {author} {\bibfnamefont
  {B.}~\bibnamefont {Wang}},\ and\ \bibinfo {author} {\bibfnamefont
  {P.}~\bibnamefont {Lu}},\ }\href@noop {} {\bibfield  {journal} {\bibinfo
  {journal} {Phys. Rev. Appl.}\ }\textbf {\bibinfo {volume} {14}},\ \bibinfo
  {pages} {064076} (\bibinfo {year} {2020})}\BibitemShut {NoStop}%
\bibitem [{\citenamefont {Xiao}\ \emph {et~al.}(2020)\citenamefont {Xiao},
  \citenamefont {Deng}, \citenamefont {Wang}, \citenamefont {Zhu},
  \citenamefont {Wang}, \citenamefont {Yi},\ and\ \citenamefont
  {Xue}}]{xiao2020non}%
  \BibitemOpen
  \bibfield  {author} {\bibinfo {author} {\bibfnamefont {L.}~\bibnamefont
  {Xiao}}, \bibinfo {author} {\bibfnamefont {T.}~\bibnamefont {Deng}}, \bibinfo
  {author} {\bibfnamefont {K.}~\bibnamefont {Wang}}, \bibinfo {author}
  {\bibfnamefont {G.}~\bibnamefont {Zhu}}, \bibinfo {author} {\bibfnamefont
  {Z.}~\bibnamefont {Wang}}, \bibinfo {author} {\bibfnamefont {W.}~\bibnamefont
  {Yi}},\ and\ \bibinfo {author} {\bibfnamefont {P.}~\bibnamefont {Xue}},\
  }\href@noop {} {\bibfield  {journal} {\bibinfo  {journal} {Nat. Phys.}\ ,\
  \bibinfo {pages} {1}} (\bibinfo {year} {2020})}\BibitemShut {NoStop}%
\bibitem [{\citenamefont {Zhang}\ and\ \citenamefont
  {Gong}(2020)}]{zhang2020non}%
  \BibitemOpen
  \bibfield  {author} {\bibinfo {author} {\bibfnamefont {X.}~\bibnamefont
  {Zhang}}\ and\ \bibinfo {author} {\bibfnamefont {J.}~\bibnamefont {Gong}},\
  }\href@noop {} {\bibfield  {journal} {\bibinfo  {journal} {Phys. Rev. B}\
  }\textbf {\bibinfo {volume} {101}},\ \bibinfo {pages} {045415} (\bibinfo
  {year} {2020})}\BibitemShut {NoStop}%
\bibitem [{\citenamefont {Gao}\ \emph {et~al.}(2020{\natexlab{b}})\citenamefont
  {Gao}, \citenamefont {Willatzen},\ and\ \citenamefont
  {Christensen}}]{gao2020anomalous}%
  \BibitemOpen
  \bibfield  {author} {\bibinfo {author} {\bibfnamefont {P.}~\bibnamefont
  {Gao}}, \bibinfo {author} {\bibfnamefont {M.}~\bibnamefont {Willatzen}},\
  and\ \bibinfo {author} {\bibfnamefont {J.}~\bibnamefont {Christensen}},\
  }\href@noop {} {\bibfield  {journal} {\bibinfo  {journal} {Phys. Rev. Lett.}\
  }\textbf {\bibinfo {volume} {125}},\ \bibinfo {pages} {206402} (\bibinfo
  {year} {2020}{\natexlab{b}})}\BibitemShut {NoStop}%
\bibitem [{\citenamefont {Wang}\ \emph {et~al.}(2021)\citenamefont {Wang},
  \citenamefont {Xu}, \citenamefont {Li}, \citenamefont {Xu}, \citenamefont
  {Chen},\ and\ \citenamefont {Wang}}]{wang2021majorana}%
  \BibitemOpen
  \bibfield  {author} {\bibinfo {author} {\bibfnamefont {Z.-H.}\ \bibnamefont
  {Wang}}, \bibinfo {author} {\bibfnamefont {F.}~\bibnamefont {Xu}}, \bibinfo
  {author} {\bibfnamefont {L.}~\bibnamefont {Li}}, \bibinfo {author}
  {\bibfnamefont {D.-H.}\ \bibnamefont {Xu}}, \bibinfo {author} {\bibfnamefont
  {W.-Q.}\ \bibnamefont {Chen}},\ and\ \bibinfo {author} {\bibfnamefont
  {B.}~\bibnamefont {Wang}},\ }\href@noop {} {\bibfield  {journal} {\bibinfo
  {journal} {Phys. Rev. B}\ }\textbf {\bibinfo {volume} {103}},\ \bibinfo
  {pages} {134507} (\bibinfo {year} {2021})}\BibitemShut {NoStop}%
\bibitem [{\citenamefont {Zhou}(2020)}]{zhou2020non}%
  \BibitemOpen
  \bibfield  {author} {\bibinfo {author} {\bibfnamefont {L.}~\bibnamefont
  {Zhou}},\ }\href@noop {} {\bibfield  {journal} {\bibinfo  {journal} {Phys.
  Rev. B}\ }\textbf {\bibinfo {volume} {101}},\ \bibinfo {pages} {014306}
  (\bibinfo {year} {2020})}\BibitemShut {NoStop}%
\bibitem [{\citenamefont {Cao}\ \emph {et~al.}(2021)\citenamefont {Cao},
  \citenamefont {Li}, \citenamefont {Chen},\ and\ \citenamefont
  {Yang}}]{cao2021universal}%
  \BibitemOpen
  \bibfield  {author} {\bibinfo {author} {\bibfnamefont {Y.}~\bibnamefont
  {Cao}}, \bibinfo {author} {\bibfnamefont {Y.}~\bibnamefont {Li}}, \bibinfo
  {author} {\bibfnamefont {Y.}~\bibnamefont {Chen}},\ and\ \bibinfo {author}
  {\bibfnamefont {X.}~\bibnamefont {Yang}},\ }\href@noop {} {\bibfield
  {journal} {\bibinfo  {journal} {ArXiv Preprint ArXiv:2103.17157}\ } (\bibinfo
  {year} {2021})}\BibitemShut {NoStop}%
\bibitem [{\citenamefont {Schomerus}(2020)}]{schomerus2020nonreciprocal}%
  \BibitemOpen
  \bibfield  {author} {\bibinfo {author} {\bibfnamefont {H.}~\bibnamefont
  {Schomerus}},\ }\href@noop {} {\bibfield  {journal} {\bibinfo  {journal}
  {Phys. Rev. Res.}\ }\textbf {\bibinfo {volume} {2}},\ \bibinfo {pages}
  {013058} (\bibinfo {year} {2020})}\BibitemShut {NoStop}%
\bibitem [{\citenamefont {Ghatak}\ \emph {et~al.}(2020)\citenamefont {Ghatak},
  \citenamefont {Brandenbourger}, \citenamefont {van Wezel},\ and\
  \citenamefont {Coulais}}]{ghatak2020observation}%
  \BibitemOpen
  \bibfield  {author} {\bibinfo {author} {\bibfnamefont {A.}~\bibnamefont
  {Ghatak}}, \bibinfo {author} {\bibfnamefont {M.}~\bibnamefont
  {Brandenbourger}}, \bibinfo {author} {\bibfnamefont {J.}~\bibnamefont {van
  Wezel}},\ and\ \bibinfo {author} {\bibfnamefont {C.}~\bibnamefont
  {Coulais}},\ }\href@noop {} {\bibfield  {journal} {\bibinfo  {journal} {Proc.
  Nat. Acad. Sci.}\ }\textbf {\bibinfo {volume} {117}},\ \bibinfo {pages}
  {29561} (\bibinfo {year} {2020})}\BibitemShut {NoStop}%
\bibitem [{\citenamefont {Longhi}(2019)}]{longhi2019probing}%
  \BibitemOpen
  \bibfield  {author} {\bibinfo {author} {\bibfnamefont {S.}~\bibnamefont
  {Longhi}},\ }\href@noop {} {\bibfield  {journal} {\bibinfo  {journal} {Phys.
  Rev. Res.}\ }\textbf {\bibinfo {volume} {1}},\ \bibinfo {pages} {023013}
  (\bibinfo {year} {2019})}\BibitemShut {NoStop}%
\bibitem [{\citenamefont {Rafi-Ul-Islam}\ \emph
  {et~al.}(2021{\natexlab{d}})\citenamefont {Rafi-Ul-Islam}, \citenamefont
  {Siu}, \citenamefont {Sahin}, \citenamefont {Lee},\ and\ \citenamefont
  {Jalil}}]{rafi2021critical}%
  \BibitemOpen
  \bibfield  {author} {\bibinfo {author} {\bibfnamefont {S.}~\bibnamefont
  {Rafi-Ul-Islam}}, \bibinfo {author} {\bibfnamefont {Z.~B.}\ \bibnamefont
  {Siu}}, \bibinfo {author} {\bibfnamefont {H.}~\bibnamefont {Sahin}}, \bibinfo
  {author} {\bibfnamefont {C.~H.}\ \bibnamefont {Lee}},\ and\ \bibinfo {author}
  {\bibfnamefont {M.}~\bibnamefont {Jalil}},\ }\href@noop {} {\bibfield
  {journal} {\bibinfo  {journal} {ArXiv Preprint ArXiv:2108.02457}\ } (\bibinfo
  {year} {2021}{\natexlab{d}})}\BibitemShut {NoStop}%
\bibitem [{\citenamefont {Okuma}\ \emph {et~al.}(2020)\citenamefont {Okuma},
  \citenamefont {Kawabata}, \citenamefont {Shiozaki},\ and\ \citenamefont
  {Sato}}]{okuma2020topological}%
  \BibitemOpen
  \bibfield  {author} {\bibinfo {author} {\bibfnamefont {N.}~\bibnamefont
  {Okuma}}, \bibinfo {author} {\bibfnamefont {K.}~\bibnamefont {Kawabata}},
  \bibinfo {author} {\bibfnamefont {K.}~\bibnamefont {Shiozaki}},\ and\
  \bibinfo {author} {\bibfnamefont {M.}~\bibnamefont {Sato}},\ }\href@noop {}
  {\bibfield  {journal} {\bibinfo  {journal} {Phys. Rev. Lett.}\ }\textbf
  {\bibinfo {volume} {124}},\ \bibinfo {pages} {086801} (\bibinfo {year}
  {2020})}\BibitemShut {NoStop}%
\bibitem [{\citenamefont {Li}\ \emph {et~al.}(2020{\natexlab{a}})\citenamefont
  {Li}, \citenamefont {Lee}, \citenamefont {Mu},\ and\ \citenamefont
  {Gong}}]{li2020critical}%
  \BibitemOpen
  \bibfield  {author} {\bibinfo {author} {\bibfnamefont {L.}~\bibnamefont
  {Li}}, \bibinfo {author} {\bibfnamefont {C.~H.}\ \bibnamefont {Lee}},
  \bibinfo {author} {\bibfnamefont {S.}~\bibnamefont {Mu}},\ and\ \bibinfo
  {author} {\bibfnamefont {J.}~\bibnamefont {Gong}},\ }\href@noop {} {\bibfield
   {journal} {\bibinfo  {journal} {Nat. Commun.}\ }\textbf {\bibinfo {volume}
  {11}},\ \bibinfo {pages} {1} (\bibinfo {year}
  {2020}{\natexlab{a}})}\BibitemShut {NoStop}%
\bibitem [{\citenamefont {Song}\ \emph {et~al.}(2019)\citenamefont {Song},
  \citenamefont {Yao},\ and\ \citenamefont {Wang}}]{song2019non}%
  \BibitemOpen
  \bibfield  {author} {\bibinfo {author} {\bibfnamefont {F.}~\bibnamefont
  {Song}}, \bibinfo {author} {\bibfnamefont {S.}~\bibnamefont {Yao}},\ and\
  \bibinfo {author} {\bibfnamefont {Z.}~\bibnamefont {Wang}},\ }\href@noop {}
  {\bibfield  {journal} {\bibinfo  {journal} {Phys. Rev. Lett.}\ }\textbf
  {\bibinfo {volume} {123}},\ \bibinfo {pages} {170401} (\bibinfo {year}
  {2019})}\BibitemShut {NoStop}%
\bibitem [{\citenamefont {Kawabata}\ \emph {et~al.}(2020)\citenamefont
  {Kawabata}, \citenamefont {Sato},\ and\ \citenamefont
  {Shiozaki}}]{kawabata2020higher}%
  \BibitemOpen
  \bibfield  {author} {\bibinfo {author} {\bibfnamefont {K.}~\bibnamefont
  {Kawabata}}, \bibinfo {author} {\bibfnamefont {M.}~\bibnamefont {Sato}},\
  and\ \bibinfo {author} {\bibfnamefont {K.}~\bibnamefont {Shiozaki}},\
  }\href@noop {} {\bibfield  {journal} {\bibinfo  {journal} {Phys. Rev. B}\
  }\textbf {\bibinfo {volume} {102}},\ \bibinfo {pages} {205118} (\bibinfo
  {year} {2020})}\BibitemShut {NoStop}%
\bibitem [{\citenamefont {Yao}\ and\ \citenamefont {Wang}(2018)}]{yao2018edge}%
  \BibitemOpen
  \bibfield  {author} {\bibinfo {author} {\bibfnamefont {S.}~\bibnamefont
  {Yao}}\ and\ \bibinfo {author} {\bibfnamefont {Z.}~\bibnamefont {Wang}},\
  }\href@noop {} {\bibfield  {journal} {\bibinfo  {journal} {Phys. Rev. Lett.}\
  }\textbf {\bibinfo {volume} {121}},\ \bibinfo {pages} {086803} (\bibinfo
  {year} {2018})}\BibitemShut {NoStop}%
\bibitem [{\citenamefont {Li}\ \emph {et~al.}(2020{\natexlab{b}})\citenamefont
  {Li}, \citenamefont {Lee},\ and\ \citenamefont {Gong}}]{li2020topological}%
  \BibitemOpen
  \bibfield  {author} {\bibinfo {author} {\bibfnamefont {L.}~\bibnamefont
  {Li}}, \bibinfo {author} {\bibfnamefont {C.~H.}\ \bibnamefont {Lee}},\ and\
  \bibinfo {author} {\bibfnamefont {J.}~\bibnamefont {Gong}},\ }\href@noop {}
  {\bibfield  {journal} {\bibinfo  {journal} {Phys. Rev. Lett.}\ }\textbf
  {\bibinfo {volume} {124}},\ \bibinfo {pages} {250402} (\bibinfo {year}
  {2020}{\natexlab{b}})}\BibitemShut {NoStop}%
\bibitem [{\citenamefont {Rafi-Ul-Islam}\ \emph {et~al.}(2022)\citenamefont
  {Rafi-Ul-Islam}, \citenamefont {Siu}, \citenamefont {Sahin}, \citenamefont
  {Lee},\ and\ \citenamefont {Jalil}}]{rafi2022critical}%
  \BibitemOpen
  \bibfield  {author} {\bibinfo {author} {\bibfnamefont {S.}~\bibnamefont
  {Rafi-Ul-Islam}}, \bibinfo {author} {\bibfnamefont {Z.~B.}\ \bibnamefont
  {Siu}}, \bibinfo {author} {\bibfnamefont {H.}~\bibnamefont {Sahin}}, \bibinfo
  {author} {\bibfnamefont {C.~H.}\ \bibnamefont {Lee}},\ and\ \bibinfo {author}
  {\bibfnamefont {M.~B.}\ \bibnamefont {Jalil}},\ }\href@noop {} {\bibfield
  {journal} {\bibinfo  {journal} {Phys. Rev. Res.}\ }\textbf {\bibinfo {volume}
  {4}},\ \bibinfo {pages} {013243} (\bibinfo {year} {2022})}\BibitemShut
  {NoStop}%
\bibitem [{\citenamefont {Li}\ \emph {et~al.}(2021{\natexlab{b}})\citenamefont
  {Li}, \citenamefont {Lee},\ and\ \citenamefont {Gong}}]{li2021impurity}%
  \BibitemOpen
  \bibfield  {author} {\bibinfo {author} {\bibfnamefont {L.}~\bibnamefont
  {Li}}, \bibinfo {author} {\bibfnamefont {C.~H.}\ \bibnamefont {Lee}},\ and\
  \bibinfo {author} {\bibfnamefont {J.}~\bibnamefont {Gong}},\ }\href@noop {}
  {\bibfield  {journal} {\bibinfo  {journal} {Commun. Phys.}\ }\textbf
  {\bibinfo {volume} {4}},\ \bibinfo {pages} {1} (\bibinfo {year}
  {2021}{\natexlab{b}})}\BibitemShut {NoStop}%
\bibitem [{\citenamefont {Zhang}\ \emph {et~al.}(2021)\citenamefont {Zhang},
  \citenamefont {Zou}, \citenamefont {Pei}, \citenamefont {He}, \citenamefont
  {Bao}, \citenamefont {Sun},\ and\ \citenamefont
  {Zhang}}]{zhang2021experimental}%
  \BibitemOpen
  \bibfield  {author} {\bibinfo {author} {\bibfnamefont {W.}~\bibnamefont
  {Zhang}}, \bibinfo {author} {\bibfnamefont {D.}~\bibnamefont {Zou}}, \bibinfo
  {author} {\bibfnamefont {Q.}~\bibnamefont {Pei}}, \bibinfo {author}
  {\bibfnamefont {W.}~\bibnamefont {He}}, \bibinfo {author} {\bibfnamefont
  {J.}~\bibnamefont {Bao}}, \bibinfo {author} {\bibfnamefont {H.}~\bibnamefont
  {Sun}},\ and\ \bibinfo {author} {\bibfnamefont {X.}~\bibnamefont {Zhang}},\
  }\href@noop {} {\bibfield  {journal} {\bibinfo  {journal} {Phys. Rev. Lett.}\
  }\textbf {\bibinfo {volume} {126}},\ \bibinfo {pages} {146802} (\bibinfo
  {year} {2021})}\BibitemShut {NoStop}%
\bibitem [{\citenamefont {Liu}\ \emph {et~al.}(2020)\citenamefont {Liu},
  \citenamefont {Zhang}, \citenamefont {Yang},\ and\ \citenamefont
  {Chen}}]{liu2020helical}%
  \BibitemOpen
  \bibfield  {author} {\bibinfo {author} {\bibfnamefont {C.-H.}\ \bibnamefont
  {Liu}}, \bibinfo {author} {\bibfnamefont {K.}~\bibnamefont {Zhang}}, \bibinfo
  {author} {\bibfnamefont {Z.}~\bibnamefont {Yang}},\ and\ \bibinfo {author}
  {\bibfnamefont {S.}~\bibnamefont {Chen}},\ }\href@noop {} {\bibfield
  {journal} {\bibinfo  {journal} {ArXiv Preprint ArXiv:2005.02617}\ } (\bibinfo
  {year} {2020})}\BibitemShut {NoStop}%
\bibitem [{\citenamefont {Sahin}\ \emph
  {et~al.}(2022{\natexlab{b}})\citenamefont {Sahin}, \citenamefont
  {Rafi-Ul-Islam}, \citenamefont {Siu}, \citenamefont {Lee},\ and\
  \citenamefont {Jalil}}]{sahin2022unconventional}%
  \BibitemOpen
  \bibfield  {author} {\bibinfo {author} {\bibfnamefont {H.}~\bibnamefont
  {Sahin}}, \bibinfo {author} {\bibfnamefont {S.}~\bibnamefont
  {Rafi-Ul-Islam}}, \bibinfo {author} {\bibfnamefont {Z.~B.}\ \bibnamefont
  {Siu}}, \bibinfo {author} {\bibfnamefont {C.~H.}\ \bibnamefont {Lee}},\ and\
  \bibinfo {author} {\bibfnamefont {M.}~\bibnamefont {Jalil}},\ }\href@noop {}
  {\bibfield  {journal} {\bibinfo  {journal} {Bull. Am. Phys. Soc.}\ }
  (\bibinfo {year} {2022}{\natexlab{b}})}\BibitemShut {NoStop}%
\bibitem [{\citenamefont {Yokomizo}\ and\ \citenamefont
  {Murakami}(2021)}]{yokomizo2021scaling}%
  \BibitemOpen
  \bibfield  {author} {\bibinfo {author} {\bibfnamefont {K.}~\bibnamefont
  {Yokomizo}}\ and\ \bibinfo {author} {\bibfnamefont {S.}~\bibnamefont
  {Murakami}},\ }\href@noop {} {\bibfield  {journal} {\bibinfo  {journal}
  {Phys. Rev. B}\ }\textbf {\bibinfo {volume} {104}},\ \bibinfo {pages}
  {165117} (\bibinfo {year} {2021})}\BibitemShut {NoStop}%
\bibitem [{\citenamefont {Kunst}\ \emph
  {et~al.}(2018{\natexlab{b}})\citenamefont {Kunst}, \citenamefont
  {Edvardsson}, \citenamefont {Budich},\ and\ \citenamefont
  {Bergholtz}}]{kunst2018biorthogonal}%
  \BibitemOpen
  \bibfield  {author} {\bibinfo {author} {\bibfnamefont {F.~K.}\ \bibnamefont
  {Kunst}}, \bibinfo {author} {\bibfnamefont {E.}~\bibnamefont {Edvardsson}},
  \bibinfo {author} {\bibfnamefont {J.~C.}\ \bibnamefont {Budich}},\ and\
  \bibinfo {author} {\bibfnamefont {E.~J.}\ \bibnamefont {Bergholtz}},\
  }\href@noop {} {\bibfield  {journal} {\bibinfo  {journal} {Phys. Rev. Lett.}\
  }\textbf {\bibinfo {volume} {121}},\ \bibinfo {pages} {026808} (\bibinfo
  {year} {2018}{\natexlab{b}})}\BibitemShut {NoStop}%
\bibitem [{\citenamefont {Lieu}(2018)}]{lieu2018topological}%
  \BibitemOpen
  \bibfield  {author} {\bibinfo {author} {\bibfnamefont {S.}~\bibnamefont
  {Lieu}},\ }\href@noop {} {\bibfield  {journal} {\bibinfo  {journal} {Phys.
  Rev. B}\ }\textbf {\bibinfo {volume} {97}},\ \bibinfo {pages} {045106}
  (\bibinfo {year} {2018})}\BibitemShut {NoStop}%
\bibitem [{\citenamefont {Lee}\ and\ \citenamefont
  {Thomale}(2019)}]{lee2019anatomy}%
  \BibitemOpen
  \bibfield  {author} {\bibinfo {author} {\bibfnamefont {C.~H.}\ \bibnamefont
  {Lee}}\ and\ \bibinfo {author} {\bibfnamefont {R.}~\bibnamefont {Thomale}},\
  }\href@noop {} {\bibfield  {journal} {\bibinfo  {journal} {Phys. Rev. B}\
  }\textbf {\bibinfo {volume} {99}},\ \bibinfo {pages} {201103} (\bibinfo
  {year} {2019})}\BibitemShut {NoStop}%
\end{thebibliography}%

\end{document}